\newif\ifsingle
\newif\ifFullVersion
\ifsingle
\documentclass[11pt,draftclsnofoot, onecolumn]{IEEEtran}		
\else		
\documentclass[10pt,final, twocolumn]{IEEEtran}
\fi

\usepackage{times}
\usepackage{amsmath,dsfont}
\usepackage{amssymb,amsthm}
\usepackage{epsfig,verbatim}
\usepackage{subfigure}
\usepackage{setspace}
\usepackage{color}
\usepackage{cite}
\usepackage{epstopdf}
\usepackage{graphics}
\usepackage{accents}
\usepackage{acronym}
\usepackage[bookmarks,colorlinks]{hyperref}
\usepackage{booktabs}
\usepackage{mathtools}
\usepackage{algorithm}
\usepackage{algorithmic}
\usepackage{enumitem}

\newtheorem{theorem}{Theorem}

\ifsingle

\else

\fi % ------------------------------------------

\acrodef{adc}[ADC]{analog-to-digital convertor}
\acrodef{aoa}[AOA]{angle of arrival}
\acrodef{coa}[COA]{curvature of arrival}
\acrodef{crb}[CRB]{Cramér Rao Bound}
\acrodef{cs}[CS]{compressed sensing}
\acrodef{csi}[CSI]{channel state information}
\acrodef{dma}[DMA]{dynamic metasurface antenna}
\acrodef{dtft}[DTFT]{discrete-time Fourier transform}
\acrodef{dnn}[DNN]{deep neural network} 
\acrodef{gps}[GPS]{global positioning system} 
\acrodef{map}[MAP]{maximum a-posteriori probability}
\acrodef{snr}[SNR]{signal-to-noise ratio}
\acrodef{sinr}[SINR]{signal-to-interference-and-noise ratio}
\acrodef{bs}[BS]{base station} 
\acrodef{em}[EM]{electromagnetic} 
\acrodef{iot}[IOT]{Interent of Things}
\acrodef{mimo}[MIMO]{multiple-input multiple-output}
\acrodef{mse}[MSE]{mean-squared error}
\acrodef{pdf}[PDF]{probability density function}
\acrodef{rv}[RV]{random variable}
\acrodef{fec}[FEC]{forward error correction}
\acrodef{lti}[LTI]{linear time-invariant}
\acrodef{mle}[MLE]{maximum likelihood estimation}
\acrodef{wss}[WSS]{wide-sense stationary}
\acrodef{psd}[PSD]{power spectral density}
\acrodef{ris}[RIS]{reconfigurable intelligent surface}
\acrodef{ser}[SER]{symbol error rate} 
\acrodef{ber}[BER]{bit error rate} 
\acrodef{sgd}[SGD]{stochastic gradient descent} 
\acrodef{toa}[TOA]{time of arrival}
\acrodef{isi}[ISI]{intersymbol interference}  
\acrodef{awgn}[AWGN]{additive white Gaussian noise} 
\acrodef{ut}[UT]{user terminal} 
\acrodef{mmw}[mmWave]{millimeter wave}
\acrodef{6g}[6G]{sixth generation}

\IEEEoverridecommandlockouts

\title{Beam Focusing for Near-Field  \\ Multi-User Localization
}
\author{  
	\IEEEauthorblockN{Qianyu Yang,~\IEEEmembership{Student,~IEEE}, Anna Guerra,~\IEEEmembership{Member,~IEEE}, Francesco Guidi,~\IEEEmembership{Member,~IEEE}, \\ Nir Shlezinger,~\IEEEmembership{Senior Member,~IEEE}, Haiyang Zhang,~\IEEEmembership{Member,~IEEE}, Davide Dardari,~\IEEEmembership{Senior Member,~IEEE},\\
Baoyun Wang,~\IEEEmembership{Senior Member,~IEEE}, and Yonina C. Eldar,~\IEEEmembership{Fellow,~IEEE}
	} 
	\thanks{
		}

}
\begin{document}
	
	\maketitle
	\pagestyle{plain}
	\thispagestyle{plain}
	%----------------------------------------------------------------------------------------
	%	ABSTRACT
	%----------------------------------------------------------------------------------------
\begin{abstract}
Extremely large-scale antenna arrays are poised to play a pivotal role in sixth-generation (6G) networks. Utilizing such arrays often results in a near-field spherical wave transmission environment, enabling the generation of focused beams, which 
%rather than conventional steering achievable in the far-field domain. Such beam focusing 
introduces new degrees of freedom for wireless localization. 
In this paper, we consider a beam-focusing design for localizing multiple sources in the radiating near-field. Our formulation accommodates various expected types of implementations of large antenna arrays, including hybrid analog/digital architectures and dynamic metasurface antennas (DMAs).
We consider a direct localization estimation method exploiting curvature-of-arrival of impinging spherical wavefront to obtain user positions. 
In this regard, we adopt a two-stage approach configuring the array to optimize near-field positioning. In the first step, we focus only on adjusting the array coefficients
%, assuming actual user positions have been known, 
to minimize the estimation error. We obtain a closed-form approximate solution based on projection and the better one based on the Riemann gradient algorithm. We then extend this approach to simultaneously localize and focus the beams via a sub-optimal iterative approach that does not rely on such knowledge.
%, which approaches the performance achievable when one has prior knowledge on where to focus the beam towards, at the cost of slightly increasing the number of iterations. 
The simulation results show that near-field localization accuracy based on a hybrid array or DMA can achieve performance close to that of fully digital arrays at a lower cost, and DMAs can attain better performance than hybrid solutions with the same aperture.
\end{abstract}

\acresetall	
\bstctlcite{IEEEexample:BSTcontrol}

%----------------------------------------------------------------------------------------
%	Introduction
%----------------------------------------------------------------------------------------

\section{Introduction}

%Radio positioning applications are expected to be notably enhanced and widely used in \ac{6g} networks~\cite{bibtex1,bibtex22,wang2022location}. The expected deployment of large antenna arrays and the usage of high frequencies will facilitate accurate radio frequency (RF) positioning, especially in \ac{gps} denied scenarios, and reduce the dependence on the global navigation satellite system for reliable localization. But these also give rise to two byproducts which have a dominant effect on the ability to carry out RF localization. First, the combination of large antennas and high frequencies implies that RF signaling likely takes place in the radiating near-field region, where the planar wavefront approximation commonly adopted in far-field systems does not hold~\cite{nepa2017near,zhang20226g}. Also, implementing arrays with a massive amount of elements is costly using conventional fully-digital designs, where each antenna is connected to a dedicated RF chain~\cite{ahmed2018survey}, so large antenna arrays utilized in 6G networks will often operate with less RF chains than elements and some degree of analog processing \cite{mendez2016hybrid}. 

In \ac{6g} networks, radio positioning is set to see significant advancements and widespread usage~\cite{bibtex1,wang2022location,torcolacci2023holographic,bellini2024multi}. The deployment of extremely large-scale antenna arrays and the utilization of high frequencies will greatly enhance radio frequency (RF) positioning accuracy, particularly in scenarios where \ac{gps} signals are inaccessible. 
However, two key challenges and opportunities emerge as dominant factors affecting RF localization. Firstly, combining extremely large-scale antennas and high frequencies often leads to RF signaling occurring within the radiating near-field region. In this region, the commonly used planar wavefront approximation, prevalent in far-field systems, no longer holds~\cite{nepa2017near,zhang20226g}.
Additionally, implementing arrays with a large number of antenna elements incurs substantial costs when using conventional fully digital designs, where each antenna necessitates a dedicated RF chain~\cite{ahmed2018survey}. Consequently, extremely large-scale antenna arrays in \ac{6g} networks often operate with fewer RF chains than elements and incorporate some degree of analog processing~\cite{mendez2016hybrid}.

Source localization can be regarded as estimating two parameters of the users, i.e., their angle and distance with respect to the reference, e.g., a base station. In far-field operating conditions, this typically becomes a joint estimation problem \cite{bibtex3, bibtex19}: obtaining the angle estimation from \ac{aoa} and obtaining the distance estimation from \ac{toa}, which requires precise synchronization and/or multiple access points participation \cite{bibtex5}. In contrast, in the radiating near field, new degrees of freedom arise from the non-negligible spherical shape of the wavefront, which can be exploited to enhance wireless localization, enabling direct localization \cite{bibtex2,DarDecGueGui:J22,9781656}. 
Additionally, unlike the TOA-based schemes, there is no need for using a wideband signal in direct localization, i.e., a user in near field can be localized using a narrowband signal. This for instance is convenient in ISAC where fewer radio resources can be dedicated to the sensing part \cite{isac}.

\begin{figure}
\centering
\includegraphics[scale=0.4]{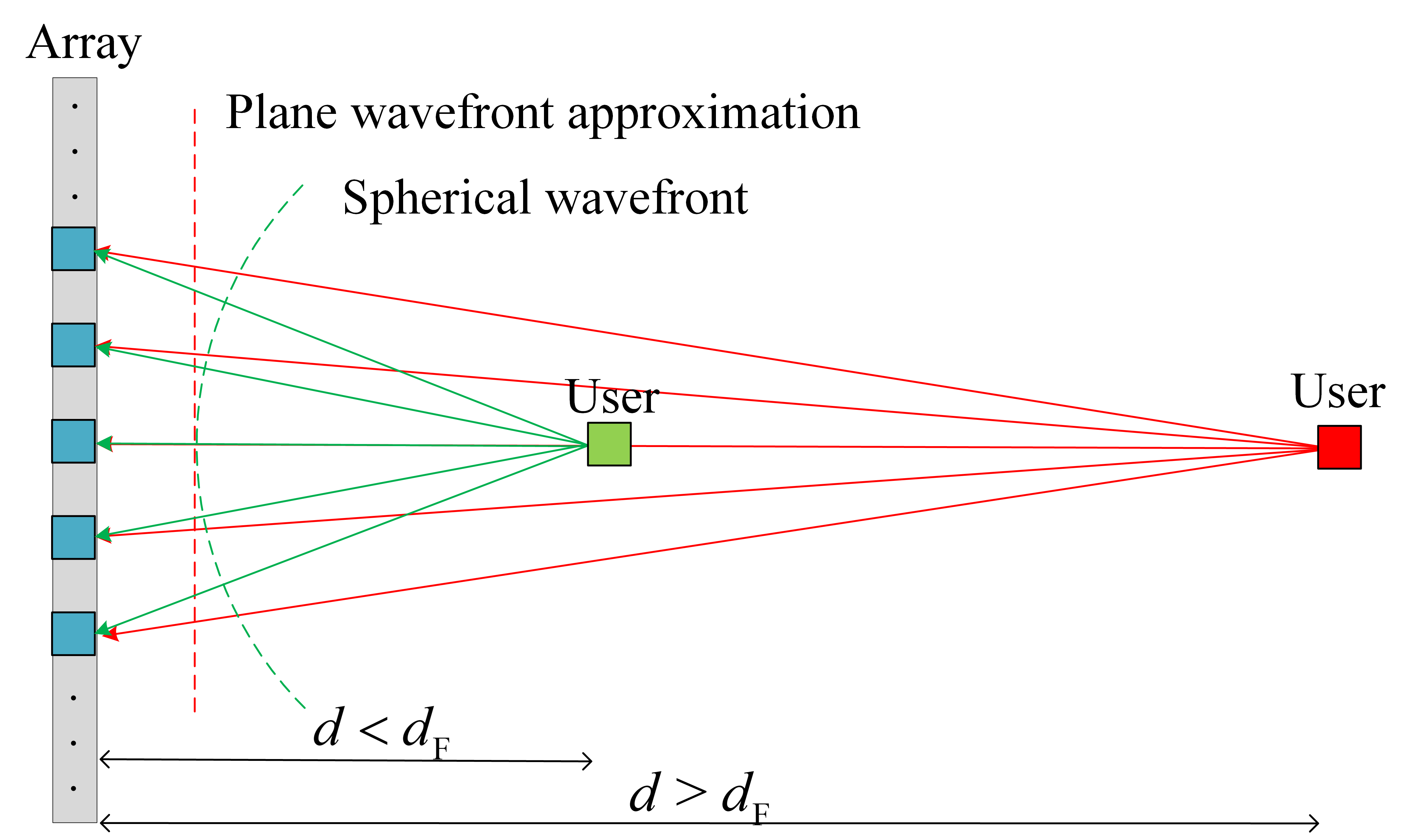}
\caption{Spherical wavefront {\em vs.} plane wavefront.
}
\vspace{0.4cm}
\label{fig}
\end{figure}

Specifically, as shown in Fig.~\ref{fig}, when the distance $d$ between the user and the array is less than the Fraunhofer distance $d_{\rm F}$, the user is considered to lie in the radiative near-field (Fresnel) region. In such cases, one can directly calculate the user's position, i.e., distance and azimuth, based on the curvature of the spherical wavefront, i.e., one can directly localize in the radiating near-field by estimating the \ac{coa}. On the contrary, when $d>d_{\rm F}$, the user is in the far-field, and the wavefront of the transmitted signal would be approximated as a plane. In this case, one can only calculate the angle using the plane waves approximation, and then obtain the distance from the delay information of the incoming signal. The mean squared error (MSE) of this two-step localization algorithm requires calculating the two-step error separately, which typically tends to achieve sub-optimal performance compared to direct localization \cite{bibtex6}.
%Therefore, near-field signal processing algorithms are typically designed to leverage the fact that the received signal has a spherical wave rather than a plane wave \cite{bibtex8}, such that one can localize in the radiating near-field by estimating the \ac{coa}.

\ac{coa}-based localization was widely used in acoustic or microwave signaling \cite{ bibtex9, bibtex10}, and it has only recently been considered for wireless communications at high-frequency bands \cite{CRB, 9950340,bibtex11, hybridRIS, HolographicRIS, bibtex13,2Dmusic,MLE}.
The recent studies of \ac{coa}-based near-field localization for \ac{6g} include the characterization of the \acl{crb} \cite{CRB,9950340}; the investigation of \acl{ris}-assisted configurations \cite{ bibtex11,hybridRIS, HolographicRIS}; the derivations of ad-hoc localization and tracking approaches \cite{bibtex13,MLE}; as well as the inclusion of hardware impairments into the performance \cite{rahal2023performance}. Several \ac{coa}-based localization works have focused on MUSIC-based processing and its simplified versions \cite{MUSIC,2Dmusic,3Dmusic}. As an example, in \cite{2Dmusic}, a reduced-dimension MUSIC algorithm for near-field sources localization is proposed to avoid the 2D spectral search (i.e., angle and distance), which has been further expanded to 3D application scenarios  \cite{3Dmusic}. 

Most works on near-field wireless positioning consider fully digital antenna architectures, focusing on the enhanced beamforming capability of the near-field. However, such antennas are expected to be too costly and power-consuming for realizing large arrays in some 6G settings \cite{zirtiloglu2022power}. 
The challenges associated with achieving fully connected extremely large-scale arrays operating at high frequencies sparked a growing interest in antenna architectures operating with reduced RF chains \cite{mendez2016hybrid}. 
The electromagnetic lens \cite{ZenZha:J16} has the potential to combat the complexity of fully digital architectures and has promoted several studies on its application in near-field localization \cite{bibtex6,abu2021near}. However, obtaining flexible and reconfigurable lenses requires complex and costly phase profile control units \cite{bibtex6}. 

An alternative solution for mitigating the complexity of fully digital architectures is to employ hybrid solutions that entail the adoption of conventional antenna technologies (e.g., based on patch arrays) while connecting multiple antennas or panels to an RF chain using dedicated analog circuitry~\cite{levy2024rapid}. Such a solution is often implemented using complex gain filter \cite{gong2019rf}, vector modulators \cite{zirtiloglu2022power}, phase-shifting networks \cite{lavi2023learn}, or switch-based operation \cite{mendez2016hybrid}. For example, \cite{hybridRIS} employs the hybrid \acl{ris} in near-field user localization and obtains improved performance of localization by optimizing the hybrid \acl{ris} configuration. 
An alternative emerging technology is based on \acp{dma},  that inherently operate with reduced RF chains without dedicated analog circuitry \cite{bibtex20}. 
DMAs have been shown to support high-rate communications with reduced RF chains in far-field \cite{shlezinger2019dynamic} and near-field \cite{bibtex15} wireless communication. 

However, the application of DMA and hybrid architectures for localization in radiating near-field has received limited attention in the literature. 
Near-field single-user location estimation for \ac{6g} systems using DMAs has been preliminary studied in \cite{DMAlocation} for a single-user scenario, and was first proposed as a beam focusing model for DMA tuning. 
On this basis, \cite{DMAfocusing} further analyzed the focus gain loss caused by the mismatch between the DMA focusing position and the actual user position, and design a non-uniform coordinate grid for effectively sampling the user area of interest.
Here, we extend the above analysis to the near-field multiple-users case, discussing the near-field localization based on the DMA and, more generally, on hybrid arrays.
%demonstrating that the conclusions we obtained in the DMA case can also be generalized to hybrid architectures.

%More specifically, 
The main contributions are summarized as follows. 
\begin{itemize}
    \item We introduce a \ac{coa} based localization method for simultaneously localizing and beam focusing in the radiating near field. We specialize our design to different antenna architectures, including fully digital arrays, hybrid arrays, and DMAs. 
    Notably, the proposed scheme allows for reducing the number of RF chains with respect to common schemes using fully digital antennas.  We illustrate how dimensionality reduction of the received signal and inherent analog processing impact the \ac{mle} of the user location. 
    \item We propose a two-stage strategy for configuring the array to enhance near-field positioning, by formulating the analog processing of DMA and hybrid antenna arrays as a form of precoding for near-field beam focusing~\cite{bibtex15}.
    This makes it possible to improve the estimation accuracy of user position by optimizing the adjustable coefficients of \acp{dma} tuning or analog phase-shifter configuration. 
    \item The first stage involves adjusting the array coefficients based on the presumed knowledge of user positions to minimize estimation errors. This approach yields two approximate solutions, one in closed form and another one that achieves better performance entailing the use of the Riemann gradient algorithm. 
    \item Then, we extend our design into a sub-optimal iterative method that does not necessitate such prior knowledge, achieving comparable performance to the optimal method albeit with a slightly higher number of iterations. Notably, in this case, the proposed joint localization and beam-focusing method is based on internal iterations carried out by the receiver, avoiding frequent remote interactions.
    \item We provide an extensive numerical analysis to demonstrate that the proposed solution can approach the performance achieved with costly fully digital antennas by setting a suitable number of iterations that gradually refine the focused beam generated.
\end{itemize}

The rest of the paper is organized as follows: Section~\ref{sec:Model} describes the considering antenna architectures and corresponding signal models. Then it introduces the COA-based multi-source localization algorithm and formulates the problem for optimization DMA tuning or hybrid phase-shifter setting. The proposed optimization algorithm is derived in Sections~\ref{sec:Solution}, while Section~\ref{sec:Sims} presents the achieved results. Finally, Section~\ref{sec:Conclusions} provides concluding remarks.
 
\emph{Notations}: Scalar variables, vectors, and matrices are represented with lower letters, lower bold letters, and capital bold letters, respectively (e.g., $x$, $\bf x$, and $\bf X$, respectively). The term $\mathbb{C}^{N \times N}$ denotes a complex space of dimension ${N \times N}$, the superscripts $({\cdot})^{ T}$ and $({\cdot})^{ H}$ denote the transpose and Hermitian transpose, respectively, and $|\cdot|$ is the absolute value operator, and $\Vert \cdot\Vert$ denotes the Frobenius norm. ${\bf l}_m$ denotes a $m$-dimensional all one vector.
%----------------------------------------------------------------------------------------
%	System Model
%----------------------------------------------------------------------------------------

\section{System Model}
\label{sec:Model}

In this section, we first introduce the considered array architectures in Section \ref{sub:DMA}, including fully digital arrays, hybrid arrays, and DMAs, and formulate the corresponding received signal model in Section \ref{sub:receive}. Then, in Section \ref{sub:model}, we present the \ac{coa}-based localization method for near-field sources and formulate the optimization problem of designing adjustable coefficients of hybrid arrays and DMAs to improve the accuracy of source position estimation.
\begin{figure}
\centering
\subfigure[Fully digital architecture]{
\includegraphics[scale=0.6]{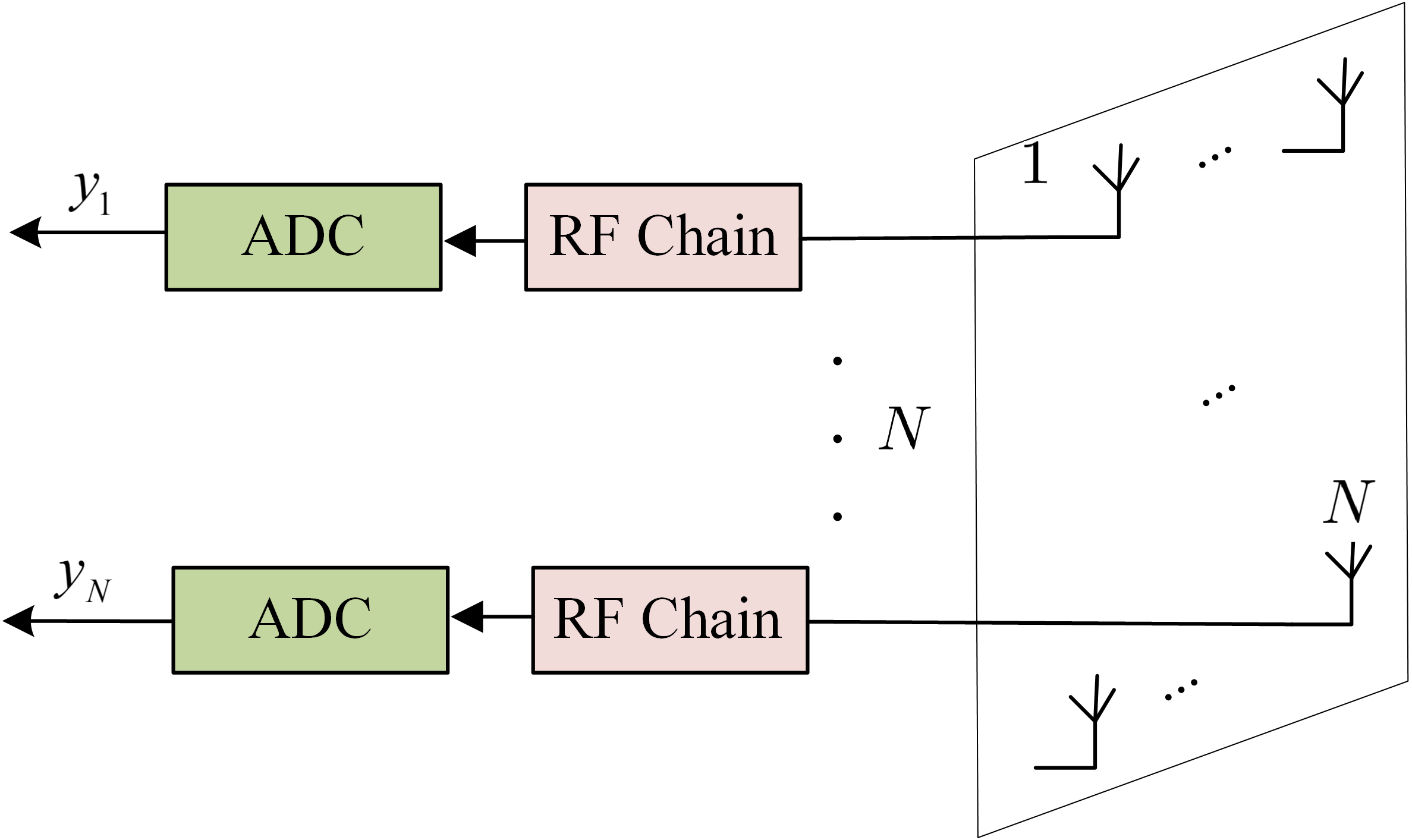}
\label{fig2.1}
}
\subfigure[Hybrid architecture]{
\includegraphics[scale=0.6]{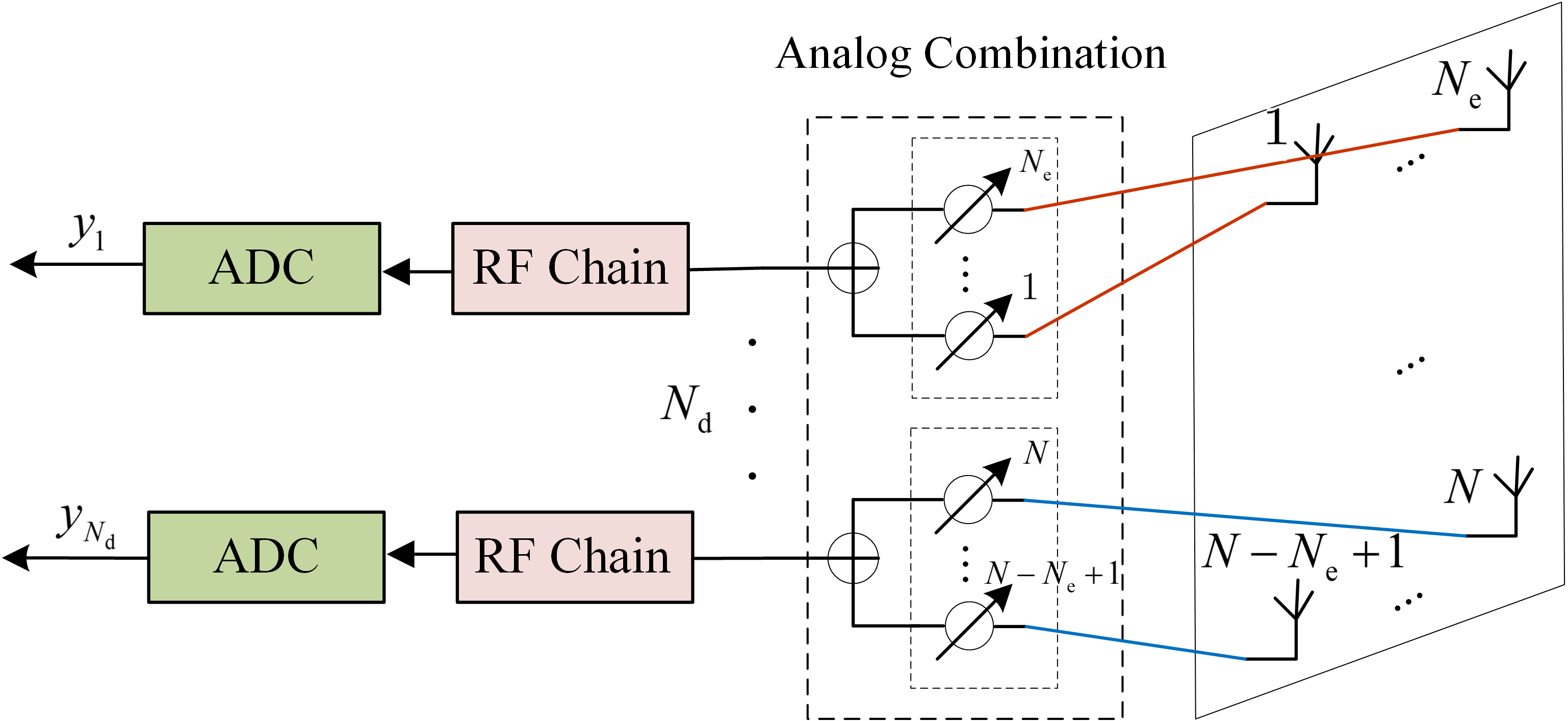 }
\label{fig2.2}
}
\subfigure[DMA architecture]{
\includegraphics[scale=0.6]{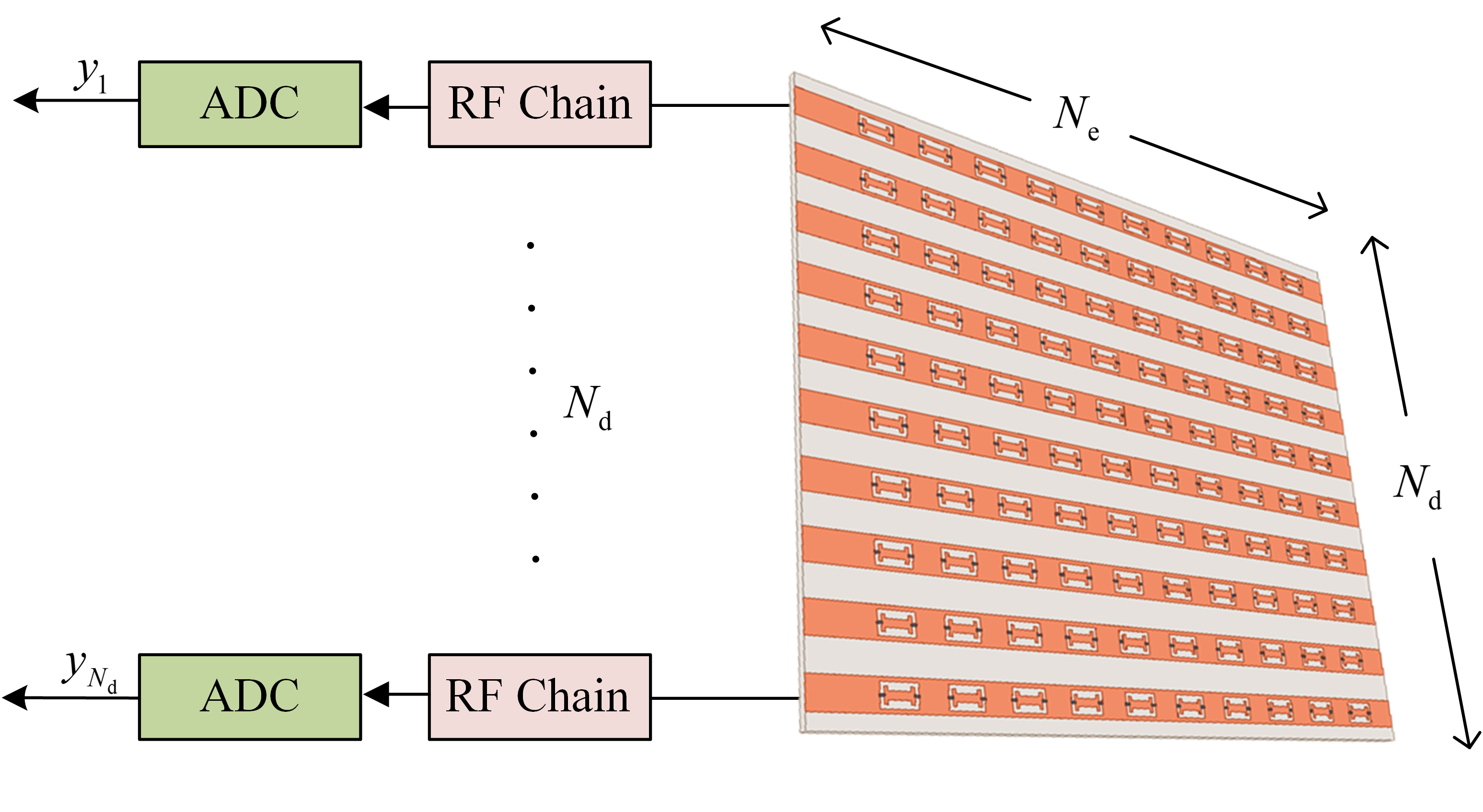}
\label{fig2.3}
}
\caption{Considered array architectures.} 
\vspace{0.4cm}
\label{fig2}
\end{figure}
\subsection{Array Architecture} 
\label{sub:DMA}

We consider three types of antenna architectures, as shown in Fig.~\ref{fig2}, that are introduced separately in the following subsections. For convenience, the antennas of the three architectures are placed in a rectangular arrangement with $N_{\rm d}$ rows and $N_{\rm e}$ columns, with an overall of $N=N_{\rm d}N_{\rm e}$ elements.

\paragraph{Fully digital architecture} 
Fully digital antennas, where processing is carried out only in digital, are widely assumed due to their high flexibility \cite{digital}.
In such architectures, each element is connected to a dedicated RF chain, as illustrated in Fig.~\ref{fig2.1}, and, thus, it results in a costly architecture when large-scale arrays are employed. In this paper, the fully digital array represents the baseline architecture to measure how much analog pre-processing (stemming from RF chain reduction) affects the localization performance.

\paragraph{Hybrid architecture}
A candidate technology to reduce the number of RF chains is represented by hybrid antenna arrays (e.g., based on patch arrays), where a large number of antennas is connected to a smaller number of RF chains using dedicated analog circuitry, such as phase-shifting networks \cite{ioushua2019family}. Phase-shifter-based hybrid arrays combine digital signal processing with some constrained level of analog signal processing. As shown in Fig.~\ref{fig2.2}, for the hybrid array, each antenna is connected to an RF chain through an independent phase-shifter, and an analog combination for antenna outputs reduces the number of RF chains. Specifically, the array is divided as $N_{\rm d}$ line sub-arrays composed of $N_{\rm e}$ elements arranged uniformly, and each line sub-array is operated by a fully connected phase-shifter network with a single RF chain. Therefore, the signal captured at the $i$th RF chain is given by
\begin{equation} 
\label{yh_i}
y_{{\rm h}i}= \sum_{l = 1}^{N_{\rm e}} { q}_{i, l} { x}_{i, l}, 
\end{equation}
where ${ x}_{i, l}$ is the received signal from the corresponding element. Here ${ q}_{i, l}$ denotes the phase of phase-shifter, which satisfies
\begin{align}
\label{lim_h}
&{ q}_{i, l} \in \mathcal{F} \triangleq\left\{e^{j \phi_{i,l}} \mid \phi_{i,l} \in[0,2 \pi]\right\}, &&\forall i, l.
\end{align}
\paragraph{DMA}
DMA is an emerging antenna technology that inherently implements RF chain reduction with low cost and power consumption and reconfigurable analog processing, using radiating metamaterial elements embedded onto the surface of a waveguide \cite{bibtex15}. 
As shown in Fig. \ref{fig2.3}, the DMA architecture comprises multiple waveguides (termed microstrips), each containing multiple metamaterial elements. The received signals of each element within each microstrip are aggregated into a corresponding RF chain \cite{bibtex20}. Since the frequency response of each element can be individually adjusted \cite{bibtex21}, the number of RF chains to be processed translates from the number of receiving elements to the number of microstrips. The elements in each microstrip are usually sub-wavelength-spaced, meaning more elements can be packed in a given aperture compared to the conventional architectures.

\begin{figure}
\centering
\includegraphics[scale=0.6]{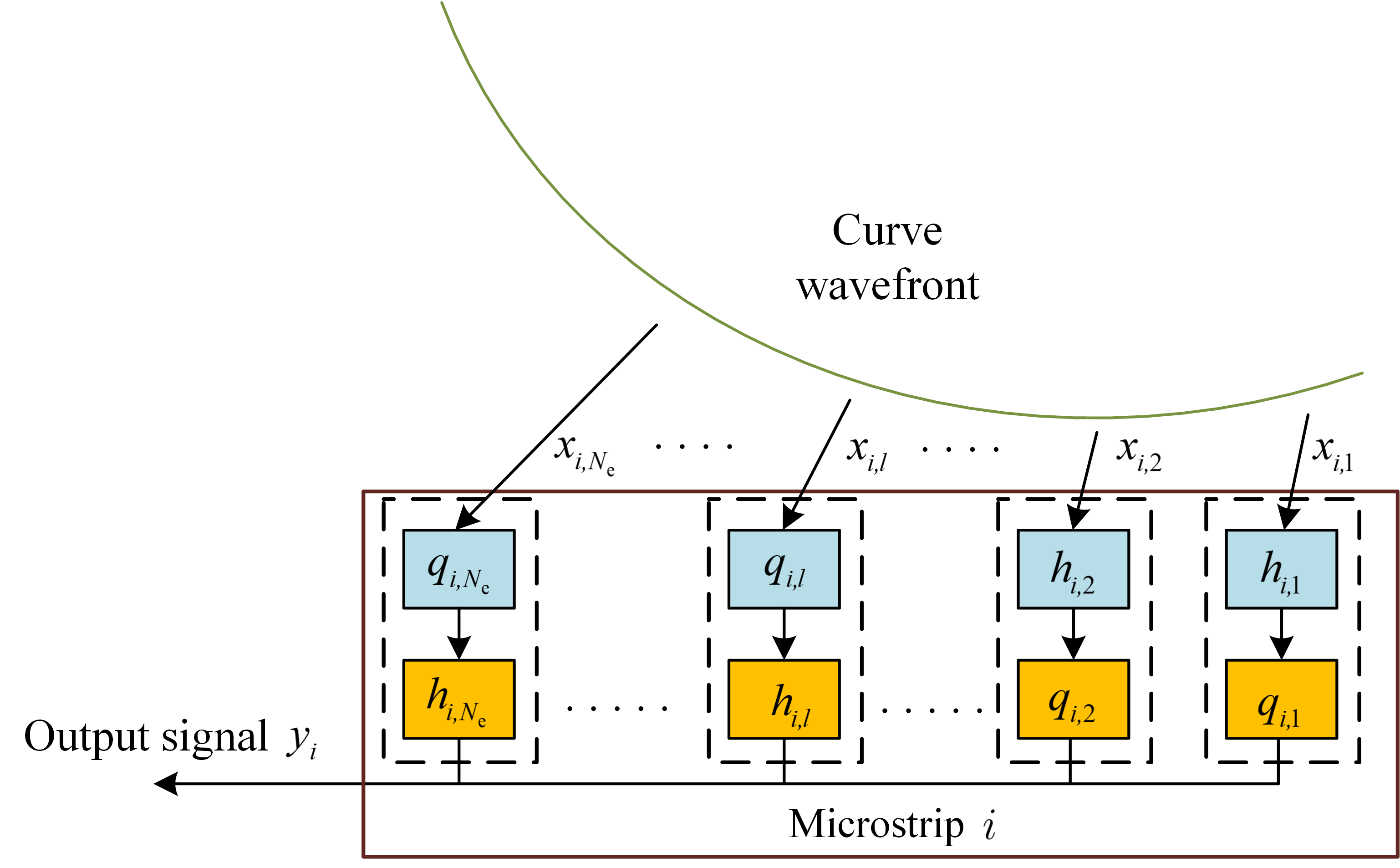}
\caption{Signal reception at the $i$-th microstrip.} 
\vspace{0.4cm}
\label{fig3}
\end{figure}

Fig.~\ref{fig3} depicts signal reception via a microstrip with multiple elements: The signals impinging on the \ac{dma} elements propagate inside the waveguide and are captured at the output port. The output of each microstrip is thus modeled as a weighted sum of these received signals; we consider the case where the response of the elements is frequency flat and focus on the Lorentzian-constrained phase model of the metamaterial elements frequency response \cite{ bibtex14}. Hence, the signal captured at the output of the $i$th microstrip, after undergoing the elements' response and propagating inside the waveguide, is given by:
\begin{equation} 
\label{y_i}
{y_i}= \sum_{l = 1}^{N_{\rm e}} { h}_{i, l}\,{ q}_{i, l} \, { x}_{i, l}, 
\end{equation}
where ${ h}_{i, l}$ encapsulates the effect of signal propagation inside the microstrip, given by \cite{shlezinger2019dynamic}
\begin{align}
\label{attenuation}
&{ h}_{i, l}=e^{-\rho_{i, l}\left(\alpha_{i}+j \beta_{i}\right)},
\end{align}
where $\alpha_{i}$ is the waveguide attenuation coefficient, $\beta_{i}$ is the wavenumber, and $\rho_{i, l}$ denotes the distance of the $l$-th element from the output port of the $i$-th microstrip. Then, the term ${ q}_{i, l}$ denotes the tunable response of the corresponding antenna, and its feasible settings satisfy the Lorentzian form
\begin{align}
\label{lim}
&{ q}_{i, l} \in \mathcal{Q} \triangleq\left\{\frac{j+e^{j \phi_{i,l}}}{2} \mid \phi_{i,l} \in[0,2 \pi]\right\}, &&\forall i, l.
\end{align}

Comparing \eqref{yh_i} and \eqref{y_i}, the received signals of both arrays are the result of accumulation, with the difference being that the weighting coefficients are the phase of phase-shifter in the hybrid array and DMA coefficient in DMA, respectively, which follows different constraints as \eqref{lim_h} and \eqref{lim}. 

\subsection{Received Signal Model}
\label{sub:receive}

We consider a multi-user localization scenario in which a multi-antenna base station receives pilot signals from $M$ users and estimates their locations. We focus on settings where the number of users is known. In practice, such information can often be reliably estimated from the impinging signals, e.g., \cite{source-numb}.
The base station is equipped with one of the three arrays of Fig.~\ref{fig2}, i.e., the common fully digital array, the hybrid array, and the DMA. As previously mentioned, although DMAs are usually distributed in a sub-wavelength space to accommodate more antennas, for the convenience of comparison, we will temporarily ignore this point in the analysis part, i.e., we keep the half-wavelength antenna space setting under the three architectures.

The array aperture and the signaling frequency are assumed to be such that the transmitting user resides in the radiating near-field, i.e., in the Fresnel region (which can be in distances of the order of tens and even hundreds of meters in some \ac{6g} settings \cite{zhang20226g}). Since the user is within the Fresnel region of the base station array, the transmitted signal exhibits a spherical wavefront.

We suppose that all users share the same pilot frequency, and the unified narrowband pilot signal at frequency $f_{\rm p}$ is emitted by each source with power normalized to one. Thus, the signal received by the $l$-th antenna of the $i$-th line array/microstrip at a generic discrete time instance is given by
\begin{align}
{ x}_{i, l} =\sum_{m=1}^{M} x_{m,i,l} +z_{i,l}.  
\label{receive}
\end{align}
In \eqref{receive}, $z_{i,l}$ is an additive Gaussian noise on the responding antenna with variance $\sigma^2$, $x_{m,i,l}$
is the received signal from the $m$-th user at the corresponding antenna. The latter can be expressed as 
\begin{align}
\begin{aligned}
{ x}_{m,i, l} & = {\rm g}_{m,i,l} \,\, x_0, 
\label{each receive}
\end{aligned}
\end{align}
where $x_0$ is the shared source pilot signal satisfies $|x_0|^2 = 1$, ${\rm g}_{m,i, l} = a_{m,i, l}  e^{-j v_{m,i, l}}$ represents the channel component, with ${ a}_{m,i, l}$ and $v_{m,i, l}$ denoting the channel gain coefficient and the phase due to the distance traveled by the signal, respectively. The gain is expressed as $a_{m,i, l}=\frac{c}{ 4\pi f_{\rm p} d_{m,i, l}}$, and the phase is given by 
\begin{equation} 
\begin{aligned}
\label{phase1}
v_{m,i, l} \triangleq  2 \pi  f_{\rm p} \frac{ d_{m,i, l}}{c},
\end{aligned}
\end{equation}
where $ d_{m,i, l}$ is the distance between the corresponding antenna and the source, and $c$ is the speed of light.

Define the vector $ {\bf x}=\left[ { x}_{1,1}, \cdots,  { x}_{i,l}, \cdots, { x}_{N_{\rm d},N_{\rm e}} \right]^T \in \mathbb{C}^{ N}$. Now, \eqref{receive} can be written as 
\begin{equation} 
\begin{aligned}
\label{x}
{\bf x} &=\sum_{m=1}^{M} {\bf g}_m \, x_0 + {\bf z} = {\bf G} \, {\bf x}_0 + {\bf z},
\end{aligned}
\end{equation}
where $ {\bf g}_m=\left[ {\rm g}_{m,1,1}, \cdots,  {\rm g}_{m,i,l}, \cdots, {\rm g}_{m,N_{\rm d},N_{\rm e}} \right]^T \in \mathbb{C}^{ N \times 1}$ and $ {\bf z}=\left[z_{1,1}, \cdots, z_{i,l}, \cdots, z_{N_{\rm d},N_{\rm e}} \right]^T \in \mathbb{C}^{N \times 1}$ denote the channel vector and noise vector, respectively. The matrix ${\bf G}=\left [ {\bf g}_1, \cdots, {\bf g}_M \right ] \in \mathbb{C}^{ N \times M}$ is the channel matrix, ${\bf x}_0 = x_0 {\bf l}_M $. Since the fully digital array assigned RF chains to all antennas, \eqref{x} can also regarded as the output of the fully digital array.

For the hybrid array, by \eqref{y_h}, the output vector ${\bf y}_{\rm h} = \left[ {y}_{{\rm h} 1}, \cdots, {y}_{{\rm h} N_{\rm d}} \right]^T \in \mathbb{C}^{ N_{\rm d}}$ is given by
\begin{equation} 
\label{y_h}
{\bf y}_h={\bf Q} \, {\bf x}, 
\end{equation}
where the matrix ${\bf Q} \in \mathbb{C}^{N_{\rm d} \times N}$ represents analog precoding of the hybrid array that satisfies 
\begin{equation}
\label{stru_h}
{\bf Q}_{n, (i-1) {N_{\rm e}}+l}= \begin{cases}{ q}_{i, l} \in \mathcal{F} & i=n, \\ 0 & i \neq n.\end{cases}
\end{equation}

Similarly, the output of DMA in vector form is expressed as
\begin{equation} 
\label{y}
{\bf y}_{\rm q}={\bf Q} \, {\bf H} \, {\bf x}, 
\end{equation}
where $\bf H$ is a $ N \times N$ diagonal matrix with diagonal elements ${\bf H}_{\left((i-1) {N_{\rm e}}+l,(i-1) {N_{\rm e}}+l\right)}={ h}_{i, l}$. The matrix ${\bf Q} \in \mathbb{C}^{N_{\rm d} \times N}$ in DMA case denotes the reconfigurable weights of the \acp{dma}.
% (it is functionally equivalent to analog precoding, so we use the same symbol to represent them).
Since we focus on narrowband signaling, for which the elements can be approximated with the Lorentzian-constrained form \cite{bibtex15}, ${\bf Q}$ obeys the following structure:
\begin{equation}
\label{stru}
{\bf Q}_{n, (i-1) {N_{\rm e}}+l}= \begin{cases}{ q}_{i, l} \in \mathcal{Q} & i=n, \\ 0 & i \neq n.\end{cases}
\end{equation}
We use the same symbol $\bf Q$ to define the adjustable coefficient matrix in \eqref{y_h} and \eqref{y}. This is because, as noted above,  DMAs and hybrid arrays perform similar operations in the analog domain (with different constraints) and yield similar output forms. Accordingly, in the following, we will also generically refer to ${\bf y}_{\rm h}$ and ${\bf y}_{\rm q}$ as $\bf y$. 

\subsection{Problem Formulation} \label{sub:model}

\begin{figure}
\centering
\includegraphics[scale=0.9]{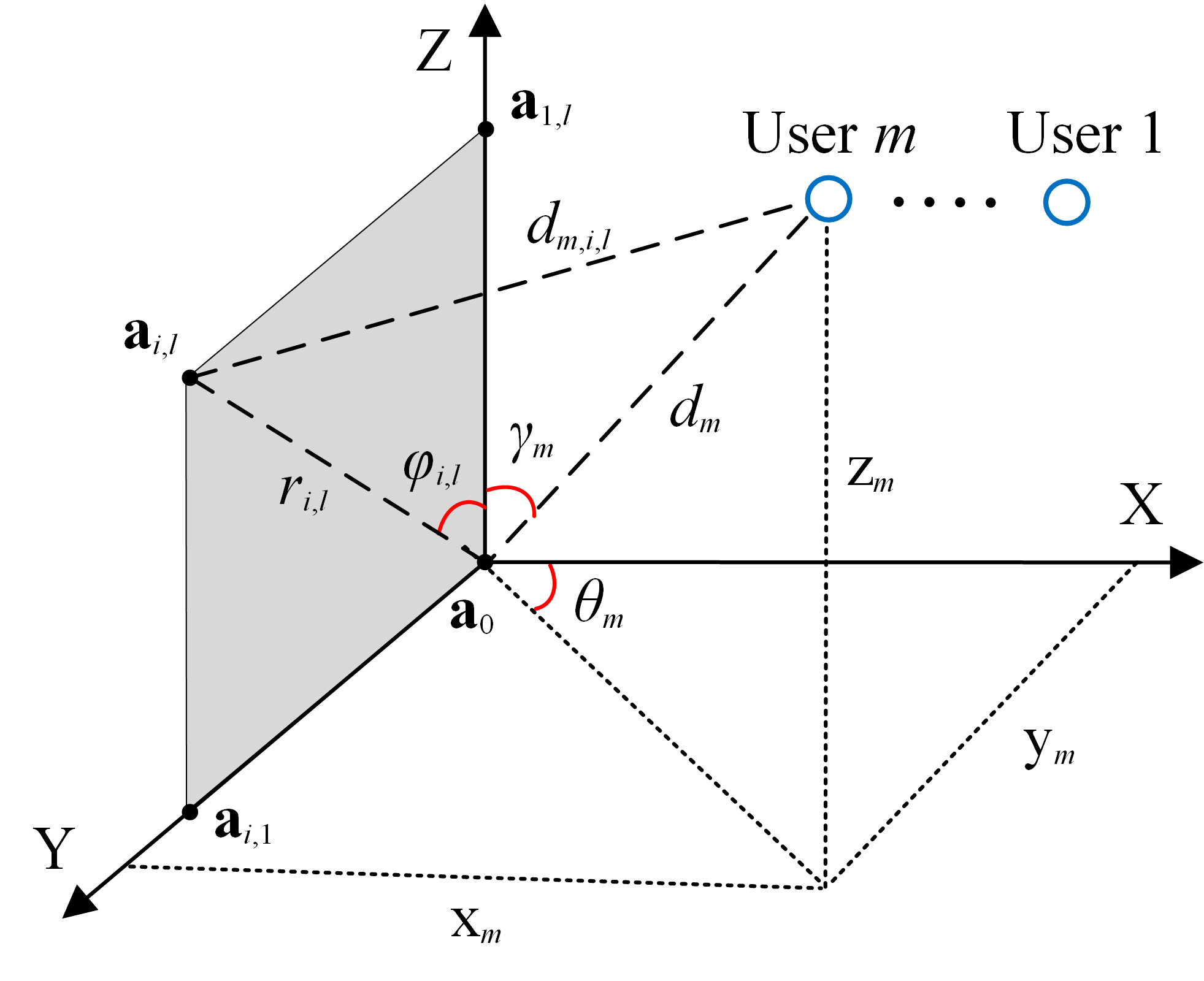}
\caption{Relative position relationships between source and array. The $\mathbf{a}_0$ is reference antenna, whereas the antenna of the $i$-line and $l$-row is at $\mathbf{a}_{i,l}$.
}
\vspace{0.4cm}
\label{fig4}
\end{figure} 

We consider the task of estimating the user positions based on the above output formulas while tuning the weights matrix ${\bf Q}$ to achieve an accurate estimate.
The fact that the communication takes place in the radiating near-field can be exploited to facilitate localization based on the COA of the impinging signal at different elements.
In our preliminary work \cite{DMAlocation}, we discussed single-user localization with DMAs, where the DMA and the user are on the same plane.

In this paper, we consider the more general localization scenario as illustrated in Fig.~\ref{fig4}, with multiple users, all three architectures detailed above, and where the square array is located on different planes with users. 

Specifically, we set a reference point (i.e., antenna) located in $(0,0,0)$ for the receiving array, and the square array is standing on the $YZ$ plane, the $m$-th user is located in $({\rm x}_m,{\rm y}_m,{\rm z}_m)$, at distance ${d}_m$, according to an azimuth $\theta_m$ and elevation angle $\gamma_m$ with respect to the source (see Fig.~\ref{fig4}), respectively. 
Further, the $l$-th antenna of the $i$-th row of the array is located at $(0,{\rm y}_{i,l},{\rm z}_{i,l})$, here we can define $\left( {r}_{i,l},\varphi_{i,l}\right)$ to indicate its position with respect to the array reference point. The specific geometric relationships between coordinates and reference positions are as follows
\begin{equation} 
\begin{aligned}
\label{position}
&{\rm x}_m = d_m \sin{\gamma_m} \cos{\theta_m}, \\
&{\rm y}_m = d_m \sin{\gamma_m} \sin{\theta_m}, \\
&{\rm z}_m = d_m \cos{\gamma_m}.
\end{aligned}
\end{equation}

According to the triangular relationship between reference points, the antenna, and the user, ${ d}_{m,i, l}\left( d_m,\theta_m, \gamma_m \right)$ can be expressed as \cite{ bibtex13}:
\begin{equation}
\label{loca1}
{ d}_{m,i, l}\left(d_m,\theta_m,\gamma_m \right)=\sqrt{{ r}_{i, l }^{2}+{ d}_{m}^{2}-2 { r}_{i, l } { d}_{m} g(\theta_m, \gamma_m)},
\end{equation}
where $g(\theta_m, \gamma_m)$ is a geometric term given by
\begin{equation}
\label{g}
g(\theta_m, \gamma_m)=\sin{\varphi_{i,l}}\sin{\theta_m}\sin{\gamma_m} + \cos{\varphi_{i,l}}\cos{\gamma_m}.
\end{equation}

The operation in the radiating near-field implies that none of the terms in \eqref{loca1} can be neglected. Thus, \eqref{loca1} determines the relationship between the user location and the phase profile, which is highly nonlinear and is retained at the above output.
Moreover, common multi-user localization methods (e.g., signal subspace techniques) are inapplicable due to the shared pilot signal.
Therefore, a possible solution for estimating the user position is to compute the MLE given by
\begin{equation}
\label{que1}
\left(\hat{d}_{\mathcal{M}}, \hat{\theta}_{\mathcal{M}}, \hat{\gamma}_{\mathcal{M}} \right)=\underset{{\bf p}_{\mathcal{M}} }{\arg \max } \log p\left({\bf y} ; {\bf p}_{\mathcal{M}} \right),
\end{equation}
where $ \log p\left({\bf y} ; {\bf p}_{\mathcal{M}} \right)$ is the log-likelihood function of $\bf y$. Here ${\bf p}_{\mathcal{M}}$ and $\left(\hat{d}_{\mathcal{M}}, \hat{\theta}_{\mathcal{M}}, \hat{\gamma}_{\mathcal{M}} \right)$ refer to the possible and the estimated user positions set in polar coordinates, respectively, with ${\bf p}_{\mathcal{M}} = \left(d_{\mathcal{M}}^{\star}, \theta_{\mathcal{M}}^{\star}, \gamma_{\mathcal{M}}^{\star} \right)$ and $\mathcal{M} = \left\{ 1, \cdots, M \right\} $. For convenience, define ${\bf p}_m = \left(d_m^{\star}, \theta_m^{\star}, \gamma_m^{\star} \right)$ as the possible position of $m$-th user, i.e., ${\bf p}_{\mathcal{M}}$ can also expressed as ${\bf p}_{\mathcal{M}} = \left \{ {\bf p}_m \right \}_{m=1}^M$.

%\textcolor{blue}{For the fully-digital array case, \eqref{que1} has already been investigated and provides comprehensive solution in several literature \cite{bibtex9,bibtex13}. However, to our knowledge, there were no relevant discussions on this based on hybrid arrays or DMA,} 
Since the received signals of both the hybrid array and the DMA also depend on the adjustable coefficient matrix $ \bf Q $, it results that $ \bf Q $ affects the accuracy of the MLE. Therefore, it is necessary to design the phase shifters or DMA tuning to facilitate localization for the hybrid array and DMA cases. Due to the similar receive mode, we will discuss DMA-based MLE in the next section and show how our approach extends to hybrid phase-shifter arrays. 

\section{Joint Localization and Beamfocusing}
\label{sec:Solution}

In this section, we explore simultaneous multi-user localization and beam focusing using large antenna architectures. As previously mentioned, we concentrate on DMAs, where beam focusing involves tuning its reconfigurable elements.
We commence by formulating an alternating projection algorithm for near-field localization for a given DMA configuration in Section \ref{sec:localDMA}. 
Then, we derive joint localization and beam focusing in two stages. First, we assume that one has prior knowledge of the users' location and develop two  DMA tuning schemes in Section \ref{sec: opt}, with the former having low complexity and the latter having higher performance. 
Then, in Section \ref{sec:alternate}, we extend these designs to also carry out localization, i.e.,  the actual case that prior knowledge of user location does not exist. This is achieved via an iterative algorithm for joint optimization position estimation and DMA tuning. Finally, in Section \ref{sec:Discussion}, we show that while our method is formulated for DMAs, it can also be naturally adapted to both hybrid phase-shifter-based architectures as well as fully digital ones after some simplification.

\subsection{Localization for Fixed DMA Tuning} \label{sec:localDMA}
We start by considering the case in which the DMA configuration, namely the matrix ${\bf Q}$, is fixed. For such setups, we propose a method for localizing the users. 

Assuming $N_{\rm T}$ samples are collected for MLE during each observation time window, the $t$-th sample is marked as ${\bf y}(t)$, $t = 1,\cdots,N_{\rm T}$.
According to \eqref{y}, once we obtain a sufficient number of receiving samples, the log-likelihood function of ${\bf y}$ can be expressed as 
\begin{equation}
\label{logp-d}
\log p\left({\bf y} ; {\bf p}_{\mathcal{M}} \right) \propto \sum_{t = 1}^{N_{\rm T}} \Vert {\mathcal{P}}\left[{\bf S} \left( {\bf p}_{\mathcal{M}}, {\bf Q} \right)\right] {\bf y}(t) \Vert^{2},
\end{equation}
where ${\mathcal{P}}$ denotes the projection operator, which, when applied to a matrix ${\bf X}$, is given by
\begin{equation*}
{\mathcal{P}}\left[ {\bf X}\right] =  {\bf X} \left[  {\bf X}^{H} {\bf X}\right]^{-1}  {\bf X}^{H}.
\end{equation*}
 Projection in \eqref{logp-d} is applied to ${\bf S} \left( {\bf p}_{\mathcal{M}}, {\bf Q} \right) = {\bf Q}{\bf H}{\bf S}_{\rm a} \left( {\bf p}_{\mathcal{M}} \right)$, with ${\bf S}_{\rm a} \left( {\bf p}_{\mathcal{M}} \right) = \left[ {\bf s}_1, \cdots, {\bf s}_M \right] \in \mathbb{C}^{N \times M}$ denoting the steering matrix of the array\footnote{In the near field, the steering matrix/vector concept loses its usual meaning as it cannot be identified as a unique steering direction. Despite that, in this paper, we still adopt this term in a wide sense.} where each steering vector is given by
\begin{equation}
\label{steering}
{\bf s }_m= \left[ {\rm a}_{m} e^{-j v_{1,1}({\bf p}_m) }, \cdots, {\rm a}_{m} e^{-j v_{m, N_{\rm d},N_{\rm e}}({\bf p}_m) } \right]^T,
\end{equation} 
where $v_{m,i,l}({\bf p}_m)$ is obtained from \eqref{phase1} and \eqref{loca1}.
Note that ${\bf s}_m = {\bf g}_m$ only if ${\bf p}_{m} = \left( d_m,\theta_m,\gamma_m \right), \forall m \in \mathcal{M}$.

The maximization of the \eqref{logp-d} is a nonlinear, multi-dimensional maximization problem, and thus, direct processing would yield a significant amount of complexity. The alternating projection (AP) maximization technique is a conceptually simple technique for multi-dimensional maximization. The technique converts the multi-dimensional maximization problem into multiple one-dimensional problems by alternating iterations. Specifically,  the projection operator ${\mathcal{P}}\left[{\bf S} \left( {\bf p}_{\mathcal{M}}, {\bf Q} \right)\right]$ can be rewritten as \cite{alternating}
\begin{equation}
\label{proj_f}
{\mathcal{P}}\left[{\bf S} \left( {\bf p}_{\mathcal{M}}, {\bf Q} \right)\right] \!=\! {\mathcal{P}}\left[{\bf S} \left( {\bf p}_{\mathcal{M}-m}, {\bf Q} \right)\right] + {\mathcal{P}}\left[\overline{\bf S} \left( {\bf p}_{m}, {\bf Q} \right)\right],
\end{equation}
where ${\bf S} \left( {\bf p}_{\mathcal{M}-m}, {\bf Q} \right)\in \mathbb{C}^{N_{\rm d} \times M-1}$ is the steering matrix based on ${\bf p}_{\mathcal{M}-m}$, while ${\bf p}_{\mathcal{M}-m}$ is the difference set of ${\bf p}_{\mathcal{M}}$ with ${\bf p}_m$, i.e., ${\bf p}_{\mathcal{M}-m}=\left \{ {\bf p}_t \right \}_{t=1,t\neq m}^M$. In \eqref{proj_f}, $\overline{\bf S} \left( {\bf p}_{m}, {\bf Q} \right) \in \mathbb{C}^{N_{\rm d} \times 1}$ denotes the residual of ${\bf Q}{\bf H}{\bf s}_m$ when projected on ${\bf S} \left( {\bf p}_{\mathcal{M}-m}, {\bf Q} \right)$, which is given by
\begin{equation}
\label{proj-f2}
\overline{\bf S} \left( {\bf p}_{m}, {\bf Q} \right) = \left( {\bf I}_ {N_{\rm d}} - {\mathcal{P}}\left[{\bf S} \left( {\bf p}_{\mathcal{M}-m}, {\bf Q} \right)\right] \right) {\bf Q}{\bf H}{\bf s }_m.
\end{equation} 

From \eqref{proj_f}, for a given $\bf Q$, once ${\bf p}_{\mathcal{M}-m}$ has been fixed, ${\mathcal{P}}\left[{\bf S} \left( {\bf p}_{\mathcal{M}}, {\bf Q} \right)\right]$ would only be determined by ${\bf p}_m$, and thus the first term of \eqref{proj_f} can be ignored. Consequently, the maximization of the \eqref{logp-d} can be iterative, at every iteration a maximization is performed with respect to a single parameter while all the other parameters are held fixed.

Therefore, by substituting \eqref{proj_f} into \eqref{logp-d}, the problem \eqref{que1} can be simplified as solving $M$ sub-problem, and the $m$-th sub-problem is expressed as
\begin{equation}
\label{logp_d2}
\log p\left({\bf y} ; {\bf p}_{m} \right) \propto \sum_{t=1}^{N_{T}} \Vert {\mathcal{P}}\left[\overline{\bf S} \left( {\bf p}_{m}, {\bf Q} \right)\right] {\bf y}(t) \Vert^{2}.
\end{equation}
Based on this formulation, one can compute \eqref{logp_d2} and iteratively localize the user by AP method when ${\bf Q}$ is fixed, the detailed process is summarized as Algorithm \ref{alg:AP}.
\begin{algorithm}
\caption{AP algorithm for multi-user localization with given DMA tuning.
%\FraCmt{Why in the algorithm do you initialize the "position mark"? It is fine for me in a numerical simulator, but not in an algorithm where you need to estimate them since they are the unknowns.} %\FraCmt{How do we know that the number of users is $M$? If so, we have prior information on the environment. Or we can say that there is a preliminary phase where we count the number of users (like a detection procedure).}
}
\label{alg:AP}
\begin{algorithmic}[1] %每行显示行号
%\renewcommand{\algorithmicrequire}{\textbf{Initialize:}} 
%\REQUIRE Position mark $\left(d_{\mathcal{M}}, \theta_{\mathcal{M}}, \gamma_{\mathcal{M}} \right)$
\renewcommand{\algorithmicrequire}{\textbf{Initialize:}} 
\REQUIRE Set $k=1$; The maximal number of iteration $ K$; ${\bf p}_{\mathcal{M}} = \{ \emptyset \}$, thus ${\mathcal{P}}\left[{\bf S} \left( {\bf p}_{\mathcal{M}}, {\bf Q} \right)\right]= {\bf 0}$.
\renewcommand{\algorithmicrequire}{\textbf{For} $m = 1:M$} \REQUIRE 
\STATE $\overline{\bf S} \left( {\bf p}_{m}, {\bf Q} \right) = \left( {\bf I}_ {N_{\rm d}} - {\mathcal{P}}\left[{\bf S} \left( {\bf p}_{\mathcal{M}}, {\bf Q} \right)\right] \right) {\bf Q}{\bf H}{\bf s }_m$.
\STATE $\left(d_m^0, \theta_m^0, \gamma_m^0 \right)=\underset{\left({\bf p}_m \right)}{\arg \max } \sum \Vert {\mathcal{P}}\left[\overline{\bf S} \left( {\bf p}_{m}, {\bf Q} \right)\right] {\bf y} \Vert^{2}$.
\STATE ${\mathcal{P}}\left[{\bf S} \left( {\bf p}_{\mathcal{M}}, {\bf Q} \right)\right] = {\mathcal{P}}\left[{\bf S} \left( {\bf p}_{\mathcal{M}}, {\bf Q} \right)\right] + {\mathcal{P}}\left[\overline{\bf S} \left( {\bf p}_{m}, {\bf Q} \right)\right] $.
\STATE ${\bf p}_{\mathcal{M}} = {\bf p}_{\mathcal{M}} \cup {\bf p}_{m} $.
\renewcommand{\algorithmicrequire}{\textbf{End}} \REQUIRE 
\STATE Obtain the initiation position ${\bf p}_{\mathcal{M}}=\left(d_{\mathcal{M}}^0, {\theta^0}_{\mathcal{M}}, \gamma_{\mathcal{M}}^0 \right)$.
\renewcommand{\algorithmicrequire}{\textbf{While} {$k \leq  K$} \textbf{do}} \REQUIRE
\renewcommand{\algorithmicrequire}{~~\textbf{For} $m = 1:M$} \REQUIRE  
    \STATE $\overline{\bf S} \left( {\bf p}_{m}, {\bf Q} \right) = \left( {\bf I}_ {N_{\rm d}} - {\mathcal{P}}\left[{\bf S} \left( {\bf p}_{\mathcal{M}-m}, {\bf Q} \right)\right] \right) {\bf Q}{\bf H}{\bf s }_m$.
    \STATE Localization: \\ ${\bf p}_m^k = \underset{\left({\bf p}_m \right)}{\arg \max } \sum_{t=1}^{N_{T}} \Vert {\mathcal{P}}\left[\overline{\bf S} \left( {\bf p}_{m}, {\bf Q} \right)\right] {\bf y}(t) \Vert^{2}$.
    \STATE Update position set: ${\bf p}_{\mathcal{M}} = {\bf p}_{\mathcal{M}-m} \cup {\bf p}_{m}^k$. 
    \STATE ${\mathcal{P}}\left[{\bf S} \left( {\bf p}_{\mathcal{M}}, {\bf Q} \right)\right] = {\mathcal{P}}\left[{\bf S} \left( {\bf p}_{\mathcal{M}-m}, {\bf Q} \right)\right] + {\mathcal{P}}\left[\overline{\bf S} \left( {\bf p}_{m}^k, {\bf Q} \right)\right] $.
\renewcommand{\algorithmicrequire}{~~\textbf{End}} \REQUIRE 
    \STATE $k=k+1$.
\renewcommand{\algorithmicrequire}{\textbf{End while}} \REQUIRE 
\renewcommand{\algorithmicrequire}{\textbf{Output:}} \REQUIRE The position estimation ${\bf p}_{\mathcal{M}}$.
\end{algorithmic}
\end{algorithm}
Notably, such an algorithm contains an initial position estimation because an accurate initialization is critical to the global convergence of the iteration.
This initial estimation starts from a single-user scenario to estimate the first user, then fixes existing estimates and continuously increases users until obtaining the last estimate. 

The computation of Algorithm \ref{alg:AP} requires a given ${\bf Q}$. Therefore, the fact that the accuracy of this estimate depends on ${\bf Q}$ allows us to use the above MLE as guidelines for tuning the DMA along with the localization task.

\subsection{DMA Tuning  for Given Positions} \label{sec: opt}
The formulation of the localization log-likelihood in the previous sections serves as a starting point for deriving our joint localization and beamfocusing method. We do this in two stages. First, we consider designing the DMA tuning to optimize the MLE~\eqref{que1}, i.e., ignoring, for now, the fact that its computation requires knowledge of the users' positions.

For ease of analysis, as in \cite{ bibtex9}, we rewrite \eqref{logp-d} as
\begin{equation}
\label{loglike}
\log p\left({\bf y} ; {\bf p}_{\mathcal{M}} \right) \propto \operatorname{tr} \left [ {\mathcal{P}}\left[{\bf S} \left( {\bf p}_{\mathcal{M}}, {\bf Q} \right)\right] {\bf R}  \right ],
\end{equation}
where ${\bf R}$ is the empirical average of ${\bf y}{\bf y}^H $ over the observed time window, which can be equivalent expressed as ${\bf R} = {\bf Q}{\bf H} \left( {\bf G}{\bf G}^H+ \sigma^2 {\bf I}_{ N} \right){\bf H}^H {\bf Q}^H$. 
%\NirCmt{is it the covariance or the empirical covariance? I guess it is the empirical one, but then it is not clear what the summation is over since there is no notion of time or time indices. If the latter is the case, the just write that it is the empirical average of ${\bf y}{\bf y}^H $ over the observed time window.}. 
For convenience, we define a function $f\left( {\bf p}_{\mathcal{M}}, {\bf Q}\right)$ as:
\begin{equation}
\label{fun_f}
f\left( {\bf p}_{\mathcal{M}}, {\bf Q}\right) = \operatorname{tr} \left [ {\mathcal{P}}\left[{\bf S} \left( {\bf p}_{\mathcal{M}}, {\bf Q} \right)\right] {\bf R}  \right ].
\end{equation}

The MLE computation can be regarded as a search process over distance and angle to maximize \eqref{fun_f}, and the maximum value of \eqref{fun_f} is expected to converge to the true user position, i.e., $\left(\hat{d}_{\mathcal{M}}, \hat{\theta}_{\mathcal{M}}, \hat{\gamma}_{\mathcal{M}} \right) = \left(d_{\mathcal{M}}, \theta_{\mathcal{M}}, \gamma_{\mathcal{M}}  \right)$. 
Therefore, we aim to design $\bf Q$ to maximize \eqref{fun_f} at the actual position, i.e., set ${\bf p}_{\mathcal{M}}^{*} = \left(d_{\mathcal{M}}, \theta_{\mathcal{M}}, \gamma_{\mathcal{M}} \right)$, and considering the following problem:

\begin{equation}
\label{max_pro}
\begin{aligned}
&\max_{\bf Q} ~f\left( {\bf p}_{\mathcal{M}}^{*}, {\bf Q}\right) \\
&~~\text{s.t.}~~ \eqref{lim}, \eqref{stru}.
\end{aligned}
\end{equation}  

Problem \eqref{max_pro} is non-convex because of the Lorentzian-constrained $\mathcal{Q}$ form in \eqref{lim}. To overcome this limitation, we introduce below two approximate solutions to relax the problem \eqref{max_pro} into the solvable convex problem to obtain an approximate solution.

1) {\em Approximate solution based on projection:}
Substituting the definition of $\bf R$ into problem \eqref{max_pro}, and using ${\bf P}_{{\bf S}_D \left( {\bf p}_{\mathcal{M}}^{\star}, {\bf Q} \right)} {\bf Q} {\bf H} {\bf G}{\bf G}^H {\bf H}^H {\bf Q}^H = {\bf Q} {\bf H} {\bf G}{\bf G}^H {\bf H}^H {\bf Q}^H$, the problem \eqref{max_pro} can be explicitly rewritten to
\begin{equation}
\label{max_prop2}
\begin{aligned}
&\max _{\bf Q} ~\operatorname{tr} \left [ {\bf Q} {\bf H} {\bf G}{\bf G}^H {\bf H}^H {\bf Q}^H \right] \\
&~~~~~~~+\operatorname{tr} \left [ {\mathcal{P}}\left[{\bf S} \left( {\bf p}_{\mathcal{M}}, {\bf Q} \right)\right] \sigma^2 {\bf Q} {\bf H}{\bf H}^H {\bf Q}^H \right] \\
&~~\text{s.t.}~~ \eqref{lim}, \eqref{stru}.
\end{aligned}
\end{equation}

The Lorentzian-constrained $\mathcal{Q}$ form in \eqref{lim} makes the problem \eqref{max_prop2} still difficult to solve. Following \cite{bibtex15}, we tackle this by relaxing the Lorentzian constraint to the phase-only weights constraint with constant amplitude and arbitrary phase, i.e. \eqref{lim_h}, the problem can be regarded as 
\begin{equation}
\label{max_prop}
\begin{aligned}
&\max _{\bf Q} ~\operatorname{tr} \left [ {\bf Q} {\bf H} {\bf G}{\bf G}^H {\bf H}^H {\bf Q}^H \right] \\ &~~\text{s.t.}~~ {\bf Q}_{n, (i-1) {N_{\rm e}}+l}= \begin{cases}{ q}_{i, l} \in \mathcal{F} & i=n. \\ 0 & i \neq n.\end{cases}
\end{aligned}
\end{equation}

The second factor of problem \eqref{max_prop2} is omitted in \eqref{max_prop} as it is a constant. This is because following the phase-only weights constraint, we have ${\bf Q} {\bf H}{\bf H}^H {\bf Q}^H = \sum_{l=1}^{l= N_{\rm e}}{h_{i,l}^2}{\bf I}_{N_{\rm d}}$  and $\operatorname{tr} \left [ {\mathcal{P}}\left[{\bf S} \left( {\bf p}_{\mathcal{M}}, {\bf Q} \right)\right] \right] = 1$.
Following the definition of $\bf G$, problem \eqref{max_prop} can be rewritten as 
\begin{equation}
\label{max_prop_2}
\begin{aligned}
&\max _{\bf Q} ~\sum_{m=1}^M \operatorname{tr} \left [ {\bf Q} {\bf H} {\bf g}_m {\bf g}_m^H {\bf H}^H {\bf Q}^H \right] \\ &~~\text{s.t.}~~ {\bf Q}_{n, (i-1) {N_{\rm e}}+l}= \begin{cases}{ q}_{i, l} \in \mathcal{F} & i=n. \\ 0 & i \neq n.\end{cases}
\end{aligned}
\end{equation}

Since $M>1$, the DMA tunable coefficients, i.e., ${ q}_{i, l}$, would not be enough to perfectly adjust for the phase of the arrived signals. But inspired by the phase design of RIS in \cite{bibtex11}, we realized that the centroid of multiple single-user optimal solutions could determine the optimal DMA tuning for the multi-user case. Thus, the solution to the relaxed problem \eqref{max_prop_2} is stated in the following:

\begin{theorem}
 \label{thm:B}
The solution to \eqref{max_prop}, denoted ${\bf Q}^*$, is obtained by setting each non-zero element to ${ q}_{i, l}^{*}=e^{j \psi_{i, l}^{*}}$, with $\psi_{i, l}^{*}= \frac{\sum_{m=1}^M v_{m,i,l}}{M}+\rho_{i, l} \beta_{i} $. 
\end{theorem}
\begin{IEEEproof}
%The lemma is obtained following \cite[Thm. 1]{bibtex15}.
The proof is given in Appendix~\ref{app:Proof0}.
\end{IEEEproof}

\begin{figure}
\centering
\includegraphics[scale=0.5]{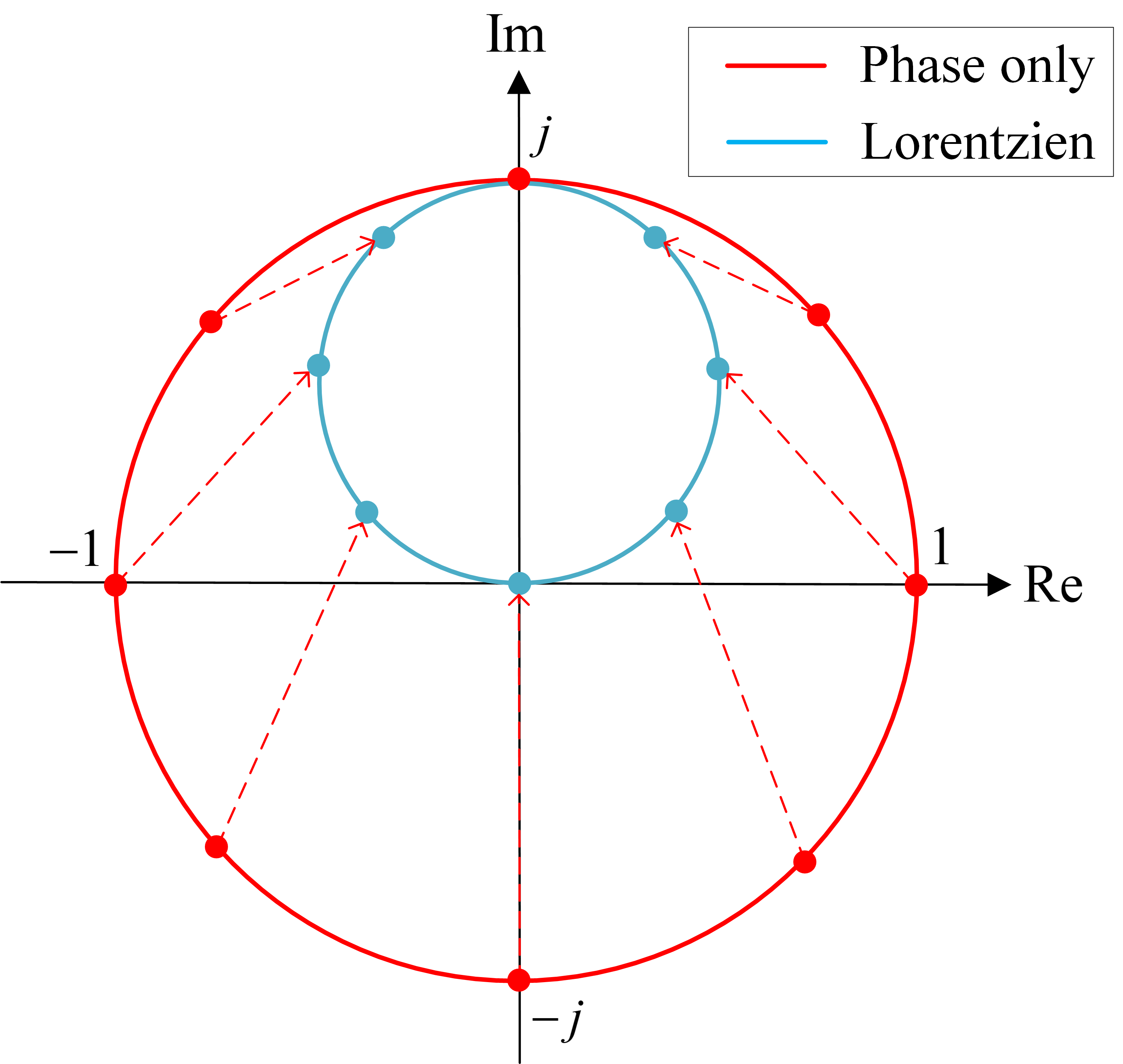}
\caption{Illustration of the phase-only weights (outer circle) and Lorentzian weights (inner circle) in the complex plane. Arrows indicate the mapping between the phase-only weights and Lorentzian weights points. }
\vspace{0.4cm}
\label{fig5}
\end{figure}

As ${ q}_{i, l}^{*}$ in Theorem~\ref{thm:B} does not satisfy the Lorentzian form, we project it onto \eqref{lim} following \cite{bibtex15}, as shown in Fig. \ref{fig5}. The resulting weight is ${ \hat q}_{i,l} = \frac{j+e^{j\psi_{i, l}^{*}}}{2}$, and though it is not the optimal solution of \eqref{max_prop2}, it provides a simple closed form solution but that guarantees reliable positioning performance as shown in the numerical results. 

In the following, we propose an alternative approach where we transform the Lorentz constraint into a phase-only constraint and use the Riemannian conjugate gradient algorithm to obtain the solution.

2) {\em Approximate solution based on the Riemannian conjugate gradient algorithm:}
To reduce the complexity of the problem, we still relax \eqref{max_prop2} as
\begin{equation}
\label{max_prop3}
\begin{aligned}
&\max _{\bf Q} ~\operatorname{tr} \left [ {\bf Q} {\bf H} {\bf G}{\bf G}^H {\bf H}^H {\bf Q}^H \right] \\ &~~\text{s.t.}~~ {\bf Q}_{n, (i-1) {N_{\rm e}}+l}= \begin{cases}{ q}_{i, l} \in \mathcal{Q} & i=n. \\ 0 & i \neq n.\end{cases}
\end{aligned}
\end{equation}

Although without the relaxation of the Lorentzian constraint $\mathcal{Q}$ makes the second factor of \eqref{max_prop2} can not be omitted, we have
\begin{equation}
\label{inequal}
\begin{aligned}
&\operatorname{tr} \left [ {\mathcal{P}}\left[{\bf S} \left( {\bf p}_{\mathcal{M}}, {\bf Q} \right)\right] \sigma^2 {\bf Q} {\bf H}{\bf H}^H {\bf Q}^H \right] \operatorname{tr} \left [ {\bf G}{\bf G}^H \right]\\
&\geq \operatorname{tr} \left [ {\bf H}^H {\bf Q}^H  {\mathcal{P}}\left[{\bf S} \left( {\bf p}_{\mathcal{M}}, {\bf Q} \right)\right] \sigma^2 {\bf Q} {\bf H}{\bf G}{\bf G}^H \right] \\
&= \sigma^2 \operatorname{tr} \left [ {\bf Q} {\bf H} {\bf G}{\bf G}^H {\bf H}^H {\bf Q}^H \right],
\end{aligned}
\end{equation}
the inequality in \eqref{inequal} use a matrix trace inequality: For any semidefinite matrix, e.g, $\bf A \succeq 0$ and $\bf B \succeq 0$, we have
\begin{equation}
\label{inequal0}
\operatorname{tr} \left[ \bf A \right]\operatorname{tr} \left[ \bf B \right] \geq \operatorname{tr} \left[ \bf AB \right],
\end{equation}
since $\operatorname{tr} \left [ {\bf G}{\bf G}^H \right]=MN$ is a constant, \eqref{inequal} demonstrates that the lower bound of the second factor of \eqref{max_prop2} can be delivered by the first factor, thus \eqref{max_prop2} can be relaxed as \eqref{max_prop3} to maximizing the lower bound of original object function.
In addition, observing \eqref{max_prop2}, it can be seen that its second factor represents the effect of DMA tuning on noise, but the localization information we are interested in is encapsulated in the signal factor, i.e., the first factor and our goal is to design the DMA tuning to improve the accurate of localization estimation. Therefore, problem \eqref{max_prop3} is sufficient for DMA tuning optimization. 
Likewise \eqref{max_prop}, \eqref{max_prop3} can be rewritten as 
\begin{equation}
\label{max_prop_3}
\begin{aligned}
&\max _{\bf Q} ~\sum_{m=1}^M \operatorname{tr} \left [ {\bf Q} {\bf H} {\bf g}_m {\bf g}_m^H {\bf H}^H {\bf Q}^H \right] \\ &~~\text{s.t.}~~ {\bf Q}_{n, (i-1) {N_{\rm e}}+l}= \begin{cases}{ q}_{i, l} \in \mathcal{Q} & i=n. \\ 0 & i \neq n.\end{cases}
\end{aligned}
\end{equation}

%The relaxation is not tight enough to make problem \eqref{max_prop_3} equivalent to \eqref{max_prop3}, however, simulation results show that the solution of \eqref{max_prop_3} is superior to the projection solution for position estimation. For convenience, in this paper, the solution of \eqref{max_prop_3} is referred to as the optimal solution and the projection solution is referred to as the sub-optimal solution.

With the Lorentzian constraint, \eqref{max_prop_3} would not obtain a closed-form solution following Theorem \ref{thm:B}. To proceed, we define the vector ${\bf q} = vec \left( {\bf Q} \right) \in \mathbb{C}^{N_{d}N \times 1}$, and thus we have ${\bf Q} {\bf H} {\bf g}_m = \left( {\bf H} {\bf g}_m \right)^T \otimes {\bf I}_{N_{\rm d}} {\bf q}$, define ${\bf W} = \left( {\bf H} {\bf g}_m \right)^T \otimes {\bf I}_{N_{\rm d}}$, \eqref{max_prop_3} can further rewritten as
\begin{equation}
\label{max_prop_4}
\begin{aligned}
&\max _{\bf q} ~\sum_{m=1}^M  {\bf q}^H {\bf W}^H {\bf W} {\bf q}  \\ 
&~~\text{s.t.}~~  { \bf q}_{n + ((i-1) {N_{\rm e}}+l-1)N_{\rm d}}= \begin{cases}{ q}_{i, l} \in \mathcal{Q} & i=n. \\ 0 & i \neq n.\end{cases}
\end{aligned}
\end{equation}

It is easy to verify that the zero elements of ${\bf q}$ do not affect the objective function of \eqref{max_prop_4}. Hence, \eqref{max_prop_4} can be simplified as
\begin{equation}
\label{max_prop_5}
\begin{aligned}
&\max _{\bf \overline{q}} ~\sum_{m=1}^M  {\bf \overline{q}}^H {\bf \overline{W}}^H {\bf \overline{W}} {\bf \overline{q}},  \\ 
&~~\text{s.t.}~~  { \overline{q}}_r \in \mathcal{Q} , \forall r \in \mathcal{A}_q,
\end{aligned}
\end{equation}
where $\mathcal{A}_q$ denotes the set of all non-zero elements of ${\bf q}$, ${\bf \overline{q}} \in \mathbb{C}^{N \times 1}$ is the modified version of ${\bf \overline{q}}$ obtained by removing all the zero elements of ${\bf q}$, and ${\bf \overline{W}}$ is the modified version of ${\bf W}$ obtained by removing the elements having the same index as the zero elements of ${\bf q}$. 

Inspired by \cite{DMApower}, to remove the Lorentzian constraint in \eqref{max_prop_5}, we define a new vector ${\bf b} \in \mathbb{C}^{N \times 1}$ whose $r$-th element is given by 
\begin{equation}
\label{b}
\begin{aligned}
b_r = 2 \overline{q}_r - e^{j 2 \pi}, \forall r \in \mathcal{A}_q.
\end{aligned}
\end{equation}
By its definition, ${\bf b}$ satisfies the phase-only constrained constant modulus, i.e., $b_r\in \mathcal{F} \triangleq\left\{e^{j \phi_{r}} \mid \phi_{r} \in[0,2 \pi]\right\}$. Meanwhile, according to \eqref{b}, we have ${\bf \overline{q}} = \frac{1}{2} \left( {\bf b} + e^{j 2 \pi} {\bf l}_N \right)$. Based on these, \eqref{max_prop_5} can be equivalently written as
\begin{equation}
\label{max_prop_6}
\begin{aligned}
&\max _{\bf b} ~\sum_{m=1}^M \frac{1}{4}  \left( {\bf b} + e^{j 2 \pi} {\bf l}_N \right)^H {\bf \overline{W}}^H {\bf \overline{W}} \left( {\bf b} + e^{j 2 \pi} {\bf l}_N \right),  \\ 
&~~\text{s.t.}~~  { {b}}_r \in \mathcal{F} , \forall r \in \mathcal{A}_q.
\end{aligned}
\end{equation}

The search space in \eqref{max_prop_6} is a product of $N$ complex circles, a Riemannian submanifold of $\mathbb{C}^N$. Therefore, \eqref{max_prop_6} can be tackled using the Riemannian conjugate gradient (RCG) algorithm. Defining the objection function of \eqref{max_prop_6} as $g\left( \bf b \right)$,
%\FraCmt{What do you mean?} 
the solution of \eqref{max_prop_6} is thus iteratively updated based on the following formula:
%\FraCmt{I am quickly looking at this part, but here I did not get why there is the time dependence. Is it only for iteration, isn't it?}
\begin{equation}
\label{b_update}
\begin{aligned}
{\bf b}^{(t+1)} = \mathcal{R}_t \left( {\bf b}^{(t)} + {\boldsymbol \epsilon}^{(t)} {\boldsymbol \eta}^{(t)} \right),
\end{aligned}
\end{equation}
where ${\bf b}^{(t+1)}$ is the updated point from the current point ${\bf b}^{(t)}$ and $t$ is the iteration number, the initial ${\bf b}^{(0)}$ could be obtained by the solution of the projection method above. Thus the RCG-based solution can be regarded as an improvement on the projection-based solution. ${\boldsymbol \epsilon}^{(t)}$ and ${\boldsymbol \eta}^{(t)}$ are the Armijo step size and the search direction at the point ${\bf b}^{(t)}$, respectively. $\mathcal{R}_t$ denotes the retraction operator that ensures the updated result satisfies constant modulus, which lets ${ b}_r^{(t+1)} = \frac{{ b}_r^{(t)} + \epsilon_r^{(t)} \eta_r^{(t)} } { \left \vert { b}^{(t)} + \epsilon_r^{(t)} \eta_r^{(t)}  \right \vert }$. 

The search direction ${\boldsymbol \eta}^{(t)}$ lies in the tangent space of the complex circle manifold at the point ${\bf b}^{(t)}$, which is given by
%\FraCmt{Has "$e$" below already been defined?}
%\FraCmt{What is $\alpha$ below? The symbol is already defined above without the $t$ index and (I think) with a different meaning.}
\begin{equation}
\label{eta}
\begin{aligned}
{\boldsymbol \eta}^{(t)}& = -\text{grad}~g\left( \bf b \right) + \\
&~~~~\zeta^{(t)} \left( {\boldsymbol \eta}^{(t-1)} - \text{Re} \left({\boldsymbol \eta}^{(t-1)} \circ {\bf b}^{(t)\dag} \right)\circ {\bf b}^{(t)} \right),
\end{aligned}
\end{equation}
where notation $\dag$ and $\circ$ conjugation denote the conjugation and Hadamard production, respectively, $\zeta^{(t)}$ is chosen as the Polak-Ribiere parameter, $\text{grad}~g\left( \bf b \right)$ is the Riemannian gradient of $ g\left( \bf b \right)$, which calculated as the orthogonally projecting of the Euclidean gradient of $ g\left( \bf b \right)$, given by
\begin{equation}
\label{grad}
\text{grad}~g\left( \bf b \right) = \nabla g\left( \bf b \right) - \text{Re} \left(\nabla g\left( \bf b \right) \circ {\bf b}^{(t)\dag} \right)\circ {\bf b}^{(t)},
\end{equation}
where $\nabla g\left( \bf b \right)$ is the Euclidean gradient of $ g\left( \bf b \right)$, which defined as
\begin{equation}
\label{deta_b}
\nabla g\left( \bf b \right) = \frac{1}{2} \left( {\bf \overline{W}}^H {\bf \overline{W}} {\bf b} + 
e^{j 2 \pi} {\bf \overline{W}}^H {\bf \overline{W}} {\bf l}_N \right).
\end{equation}

\subsection{Alternating Localization and DMA Tuning} \label{sec:alternate}

The DMA tuning detailed above requires knowledge of the user position, whose estimation is the core of the localization problem. However, alternating optimization can extend it for joint DMA tuning and localization. The rationale here stems from the fact that the design of  $\bf Q$ for \ac{coa}-based localization realizes a form of beamforming, which in the radiating near-field specializes in beamfocusing \cite{nepa2017near}. Consequently, \eqref{max_pro} can be regarded as designing $\bf Q$ to focus the received beam towards a given location and iteratively refine the specific location via the \ac{mle}.  

Therefore, for the \ac{mle} process in \eqref{loglike}, even when the focusing position at the array receiving end deviates from the actual source position, the estimated results are still improved \footnote{As we analysed, the improvement which yield from DMA tuning is caused by the focusing operation. Even if there is a mismatch between the focusing position and the actual user position, the focusing gain still exists while is only loss \cite{DMAfocusing}.}, which means that once we obtain a rough estimation of the source location, we can constantly approach the desired actual source position by updating the focus position. Meanwhile, the update of $\bf Q$ can be synchronized with the iteration in the AP algorithm, i.e., we can obtain a new alternating algorithm by embedding the update of $\bf Q$ into Algorithm \ref{alg:AP}.

\begin{algorithm}
\caption{Iterative optimization algorithm for alternating localization and DMA tuning}
\label{alg:Alternating}
\begin{algorithmic}[1] %每行显示行号
\renewcommand{\algorithmicrequire}{\textbf{Initialize:}} 
\REQUIRE Initialize the matrix ${\bf Q}^1 \in \mathbb{C}^{N_{\rm d} \times N}$ constrained by the array architecture; $k=1$;  Iterations limit $K$.
\STATE Obtain the  position estimation ${\bf p}_{\mathcal{M}}$ by Algorithm \ref{alg:AP} (Steps 1-5).
\renewcommand{\algorithmicrequire}{\textbf{While} {$k \leq  K$} \textbf{do}} \REQUIRE  
\renewcommand{\algorithmicrequire}{~~\textbf{For} $m = 1:M$} \REQUIRE 
    \STATE Estimate ${\bf p}_{m}^k$ by maximizing $f\left( {\bf p}_{\mathcal{M}}, {\bf Q}^k \right)$ (equivalently maximize the projection MLE defined in \eqref{logp_d2}).
    \STATE Update position set: ${\bf p}_{\mathcal{M}} = {\bf p}_{\mathcal{M}-m} \cup {\bf p}_{m}^k$.
\renewcommand{\algorithmicrequire}{~~\textbf{End}} \REQUIRE     
    \STATE Get ${\bf Q}^{k+1}$ by maximizing $f\left( {\bf p}_{\mathcal{M}}, {\bf Q}\right)$ through  solving \eqref{max_pro}.
    \STATE Update receive vector ${\bf y}^{k+1}$ from ${\bf Q}^{k+1}$ by \eqref{y}.
    \STATE $k=k+1$
\renewcommand{\algorithmicrequire}{\textbf{End while}} \REQUIRE
\renewcommand{\algorithmicrequire}{\textbf{Output:}} \REQUIRE The position estimation ${\bf p}_{\mathcal{M}}$.
\end{algorithmic}
\end{algorithm}

The resulting alternating localization and DMA tuning algorithm is summarized as Algorithm~\ref{alg:Alternating}. 
In Algorithm~\ref{alg:Alternating}, we utilize the receiving vector from random coefficient response performing the initial position estimation and obtain new observations after updating the coefficients of $\bf Q$ with the results of the previous position estimation. We can repeat the above steps at each receiving time slot so that the estimated result is close to the actual source location gradually. 

Compared to the Algorithm \ref{alg:AP}, the additional update of $\bf Q$ introduced in Algorithm~\ref{alg:Alternating} results in more observation time requirements, as each update of $\bf Q$ requires re-observation of the signal. However, it should be emphasized that the application of DMA will significantly reduce the demand for RF chains, as shown in numerical evaluations, which not only achieves the cost advantage of the DMA but also significantly reduces the required signal processing dimension and data processing time, which compensate for the increased observation time.

% ------------------- Discussion ------------------------

\subsection{Extension to Fully Digital and Hybrid Architectures}
\label{sec:Discussion}
In this part, we extend the proposed localization and tuning scheme to other antenna architectures.

1) {\em Fully Digital Antenna:} For the fully digital array, each antenna element is connected to a dedicated RF chain (without analog combining operation), leading to the received signal ${\bf y}={\bf x}$. In this case, the likelihood function in \eqref{que1} of $\bf x$, can be expressed as \cite{ bibtex9}:
\begin{equation}
\label{logp-f}
\log p\left({\bf x} ; {\bf p}_{\mathcal{M}} \right) \propto \sum_{t=1}^{N_{T}} \Vert {\mathcal{P}}\left[{\bf S}_{\rm a} \left( {\bf p}_{\mathcal{M}} \right)\right] {\bf x}(t) \Vert^{2},
\end{equation} 

Since without analog domain design, by making ${\bf Q}={\bf I}_N$ and ${\bf H}={\bf I}_N$, we can directly use Algorithm~\ref{alg:AP} to implement localization.

2) {\em Hybrid Phase-Shifter Antenna:} When we consider the hybrid arrays, as previously introduced, the optimization of the phase shifters is similar to DMA tuning. Thus, Algorithm~\ref{alg:AP} also applies to the hybrid array case and just needs to let ${\bf H}={\bf I}_N$.
Besides, by replacing the constraints of matrix $\bf Q$ to \eqref{lim_h} and \eqref{stru_h}, \eqref{max_pro} can be regarded as optimizing the phase of the hybrid array to improve the MLE, which can be written as
\begin{equation}
\label{max_prop_h}
\begin{aligned}
&\max _{\bf Q} ~\operatorname{tr} \left [ {\bf Q} {\bf G}{\bf G}^H {\bf Q}^H \right] \\ &~~\text{s.t.}~~ {\bf Q}_{n, (i-1) {N_{\rm e}}+l}= \begin{cases}{ q}_{i, l} \in \mathcal{F} & i=n. \\ 0 & i \neq n.\end{cases}
\end{aligned}
\end{equation}
Problem \eqref{max_prop_h} is almost identical to \eqref{max_prop}, except for the omission of the matrix $\bf H$. Thus it can obtain the solution from Theorem~\ref{thm:B} directly by removing the effect of $\bf H$. The resulting phase is ${\hat{q} }_{i, l}=e^{j \psi_{i, l}^{*}}$, with $\psi_{i, l}^{*}= \frac{\sum_{m=1}^M v_{m,i,l}}{M} $.

Since without the requirement of projection of the phase solution onto Lorentzian constraint, the solution of \eqref{max_prop_h} can be regarded as the optimal solution of the hybrid array.
In fact, the Lorentzian constraint only has a negative effect on phase design. Therefore, hybrid arrays always outperform DMA under the same number of antennas and array aperture.
 
%-----------------------------------------------------------------------------------%	NUMERICAL EVALUATIONS
%-----------------------------------------------------------------------------------

\section{Numerical Evaluations}
\label{sec:Sims}
This section evaluates the localization performance under multiple array architectures for different SNR scenarios. We first detail the simulation setup in Section~\ref{parameter}, and present our numerical results in Section~\ref{ssec:results}

\subsection{Simulation Parameters} \label{parameter}
In our simulation study, we set the carrier frequency of the pilot signal as $f_{\rm p} = 28$ GHz, corresponding to a signal wavelength of $\lambda = 0.01$ m. In one observed time window, the sample number is set as $N_{T} = 500$. 
Then, we consider a localization scenario in the $XY$-plane, while the antenna lies on the $YZ$-plane like in Fig.~\ref{fig4}. In this case, the position of each user can be expressed as $({\rm x}_m,{\rm y}_m,0)$ in Euclidean coordinate, or $(d_m, \theta_m, \pi/2)$ in polar coordinates.
%\FraCmt{In the following footnotes I will try to highlight some considerations that are missing from the numerical results.}
%\FraCmt{1. Why I should use the proposed multi-user approach rather than any single-user localization approach that is applied in temporal sequence (e.g., I first localize user 1, then user 2, then user 3, etc.)? Do we have a gain in terms of latency and/or performance? If so, we should clearly state it.}
%\FraCmt{2. Fig. 6 should be repeated with the receiver orientation of 45 degrees, such that the plot it roughly symmetric with respect to the diagonal of the room.}
%\FraCmt{Fig. 7: please write RMSE (m). Which SNR are you considering? the SNR at the single antenna or at the overall receiver? Because here we are saying that the hybrid array allows to attain RMSE of about 25 cm at 6 m for an SNR of -15 dB, which is very surprising!}
%\FraCmt{Fig. 9: please write RMSE (m). Moreover, can you provide also some insights about having two users at the same angle but at different distances as it is more of interest for near-field?}
%\FraCmt{Fig. 10: we must clarify that the top figure is in the near-field and the bottom figure is in the far-field. But I still do not understand why there is angular ambiguity in the likelihood function, since we should be in a condition like traditional far-field where we are able to distinguish users in the angular domain.}
%\FraCmt{I feel that Fig. 10 could be enriched with other sub-plots, where we can see two users at the same angle from the receiver but at different distances, like in our beamfocusing paper.}
We simulate DMAs as well as hybrid phase-shifter antennas and fully digital ones. For DMAs, we set use $\alpha = 0.6\, {\rm m}^{-1}$ and $\beta = 827.67\, {\rm m}^{-1}$ to represent the propagation inside the DMA waveguides. The DMA array is comprised of $N_{\rm d }$ microstrips, i.e., it only requires $N_{\rm d }$ RF chains, with each microstrip containing $N_{\rm e }$ antenna elements. The total number of antenna elements is thus $N = N_{\rm d }N_{\rm e }$. The noise variance $\sigma^2$ is set to adjust the signal-to-noise ratio (SNR) as $\text{SNR} = 10 \text{log}(\Vert{\bf x}\Vert^2/(N\sigma^2))$.

The separation between microstrips is set as $5\lambda/2$, (the separation between rows of hybrid or fully digital arrays is also set in this way). In general, the antenna elements within the microstrip of DMAs are arranged at sub-wavelength spacing, while the antennas of full digital arrays and hybrid arrays require at least half-wavelength spacing to be arranged. To compare the performance differences of three antenna architectures in detail, we set two kinds of DMA microstrip antenna space, one set to $\lambda /2$ antenna spacing (as in the full digital case and hybrid case), and the other set to $\lambda /4 $ antenna spacing (as in the general DMA case), the full digital array and hybrid array have the same dimensions as the DMA at the $\lambda /2 $ spacing case. The number of antennas and required RF chains for each array are summarized in Table \ref{table:1}. To make a fair comparison, we consider that the three different types of antenna architectures have the same physical aperture equal to $D = \sqrt{2}L$, with $L$ denoting the length of the microstrip. We fix $L = 25$ cm thereby the number of microstrips in this DMA is $N_{\rm d} = 2L/(5\lambda) = 10$. For $\lambda /2 $ spacing (antennas in each row of the full digital array is the same), the number of antenna elements of each microstrip is calculated as $N_{\rm e} = 2L/\lambda = 50$ and the total number of antenna elements is thus $N_1 = N_{\rm d }N_{\rm e } = 500$; For $\lambda /4 $ spacing, $N_{\rm e} = 4L/\lambda = 100$ and thus $N_2 = 1000$. 

\begin{table}[t!]
    \centering
    \caption{Array architecture.}
    \label{table:1}
    \begin{tabular}{|c|c|c|}
    \hline \textbf{Architecture} & \textbf{Antenna number} & \textbf{RF number} \\
    \hline \text{Full digital} $\lambda/2$ & $N_1 $  & $N_1$ \\
    \hline \text{DMA} $\lambda/2$ & $N_1$ & $N_{\rm d}$ \\ 
    \hline \text{Hybrid} $\lambda/2$ & $N_1$ & $N_{\rm d}$ \\ 
    \hline \text{DMA} $\lambda/4$ & $N_2$ & $N_{\rm d}$ \\ 
    \hline
    \end{tabular}
\end{table}
%\subsection{Near Field Range of DMA\FraCmt{What does this title mean? Can we also keep the analysis as general as possible, without specifying that everything is only for DMA?}}

\begin{figure}
\centering 
\includegraphics[width=1\columnwidth]{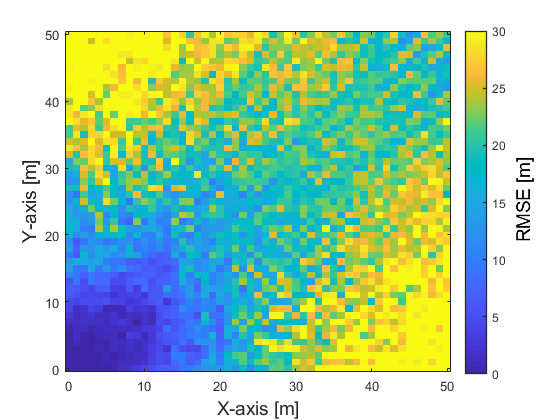}
\caption{The heatmap for position estimate RMSE under different user positions, with $N_{\rm d } = 10$, SNR is fixed as -10 dB. The RMSE significantly decreases in the Fresnel near-field region of the array.
%\textcolor{red}{Why it is not symmetric? which is the receiver orientation? I am afraid that it looks towards the $x$-axis direction.} %\FraCmt{I did not get the meaning of this figure. What do we want to show? Like this it is not easily readable.}
}
\vspace{0.4cm}
\label{fig6}
\end{figure}

\subsection{Numerical Results
%\FraCmt{In this section, I feel that a simple figure related to communication performance is missing, since beam focusing is typically conceived for multi-user communication. Can we show something related to the impact of localization error into communication performance? If not possible, we could just discuss about it with few sentences.}
}
\label{ssec:results}

The COA-based localization method is only applicable in the near field range, thus the source position $(d_m, \theta_m, \pi/2)$ should be located in the Fresnel region of the array, i.e., $d_m$ should satisfy that 
\begin{equation}
d_0 \leq {d_{\rm F} =} \frac{2 D^{2} }{ \lambda},
\end{equation}
where $d_{\rm F}$ is the Fraunhofer distance limit, under the current setting, equals $d_{\rm F} =24 $ meters. Besides, the near-field range is also affected by the array shape. 

To demonstrate this point, we provide a root mean square error (RMSE) heatmap to clarify the range of the proposed COA-based localization effectiveness, as shown in Fig.~\ref{fig6}. We use the DMA as the example to localize one user (the relevant conclusions arising from this are also applicable to other antenna architectures), the DMA is $\lambda/2$ spacing. We consider a $50\times 50 ~ {\rm m}^2$ grid of points with a resolution of 1 meter, with the DMA located in the origin. Then, we set each grid point as the target user location to estimate the user position using Algorithm~\ref{alg:Alternating} with optimal DMA tuning design. The RMSE of each estimation is calculated as
\begin{equation}
\label{rmse}
\mathsf{RMSE}= \sqrt{\frac{1}{N_{\text{mc}}} \sum_{n=1}^{N_{\text{mc}}}  e_{n}^2},
\end{equation}
where $N_{\text{mc}}$ is the number of Monte Carlo iterations, $e_{n}^2=  |\left( {\rm x}, {\rm y} \right) - \left( \hat{\rm x}^{(n)}, \hat{\rm y}^{(n)} \right)|^{2}$ is the squared localization error (i.e., the distortion), $\left( {\rm x}, {\rm y} \right)$ is the source coordinates, and $\left( \hat{\rm x}^{(n)}, \hat{\rm y}^{(n)} \right)$ are the estimated coordinates at the $n$-th Monte Carlo simulation. For the considered scenario, following the results reported in Fig.~\ref{fig6}, the RMSE of estimations would be significantly reduced within a distance of 24 meters, i.e. within the near field range.

\begin{figure}
\centering 
\includegraphics[width=1\columnwidth]{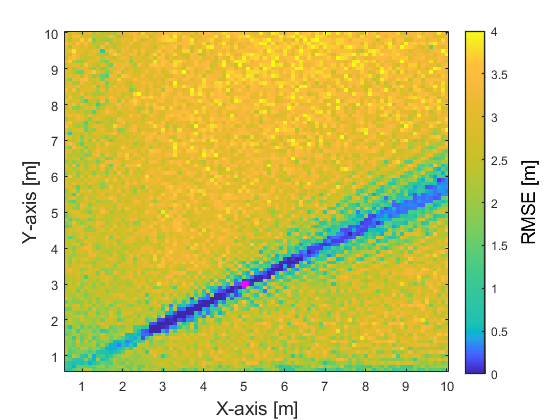}
\caption{The heatmap for the RMSE on user position estimate by varying user positions, SNR is fixed as -10 dB. The red point is the hypothetical user position that $\bf Q$ focuses on, located at $d = 0.25 \, d_{\rm F}$ meters, $\theta = \pi /6$.
}
  \vspace{0.4cm}
\label{fig7}
\end{figure}

\begin{figure}
\centering 
\includegraphics[width=1\columnwidth]{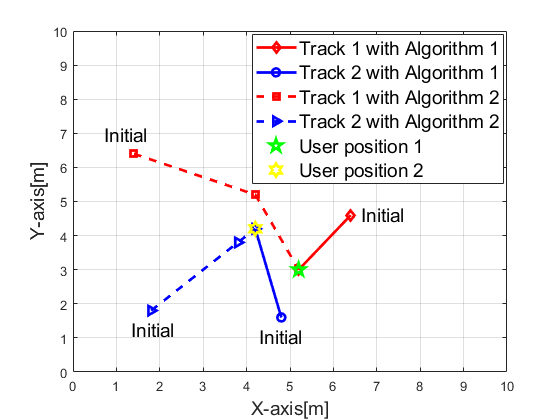}
\caption{
Position estimations track map for each iteration of Algorithm~\ref{alg:AP} and Algorithm~\ref{alg:Alternating}, SNR is fixed as -5 dB, with user position: $d_1 = 0.25 \, d_{\rm F}$ meters, $\theta_1 = \pi /6$; $d_2 = 0.25 \, d_{\rm F}$ meters, $\theta_2 = \pi /4$. 
}
  \vspace{0.4cm}
\label{fig8}
\end{figure}

To present in detail that our design of $\bf Q$ realizes a form of beam focusing, a heatmap is provided in Fig.~\ref{fig7}. We consider a $10\times 10 ~ {\rm m}^2$ grid of points with the resolution of $0.1$ meters, with the array located on the origin. Then, we set the user to transmit the pilot signal on each grid point, and the $\bf Q$ is designed to focus on the fixed point as shown (i.e., the $\bf Q$ is obtained by assuming the user is located on the setting fixed point and transmit the pilot signal). The DMA would localize the user based on the actual received pilot signal with the fixed $\bf Q$. 
We can see that the RMSE would be significantly improved only when the user position is set near the focus point of $\bf Q$, which proves our view. 
Then, we further demonstrate the effectiveness of the proposed localization algorithm, as shown in Fig.~\ref{fig8}. In Algorithm~\ref{alg:AP}, the localization is performed with a given $\bf Q$ by the proposed RCG solution. Even if it is not realistic because the actual user position is unknown, we assume that the given $\bf Q$ has been designed perfectly to achieve a fair comparison with Algorithm~\ref{alg:Alternating}). Comparatively, in Algorithm~\ref{alg:Alternating}, the localization and the $\bf Q$ design are carried out alternately. It can be seen that both track of the position estimate with two algorithms gradually approaches the real user position from the initial position estimate with iteration. Besides, it should be noted that Algorithm~\ref{alg:AP} yields a more accurate initial position estimation compared to Algorithm~\ref{alg:Alternating}, and the estimation iterates faster to the actual user position, which indicates the impact of the $\bf Q$ design on localization performance.

%the track of the position estimate from Algorithm~\ref{alg:AP} and Algorithm~\ref{alg:Alternating} gradually approaches the real user position with iteration, even if the DMA tuning design in Algorithm~\ref{alg:Alternating} does not focus on the actual user position. $\bf Q$ in Algorithm~\ref{alg:AP} is given by the proposed RCG solution (This is not realistic because the user position is unknown, but for comparison with Algorithm~\ref{alg:Alternating}, we still provide designed $\bf Q$ for Algorithm~\ref{alg:AP}), thus the track of the position estimate from Algorithm~\ref{alg:AP} obtain a more accurate initial position estimation.

\begin{figure} 
\centering 
\includegraphics[width=1\columnwidth]{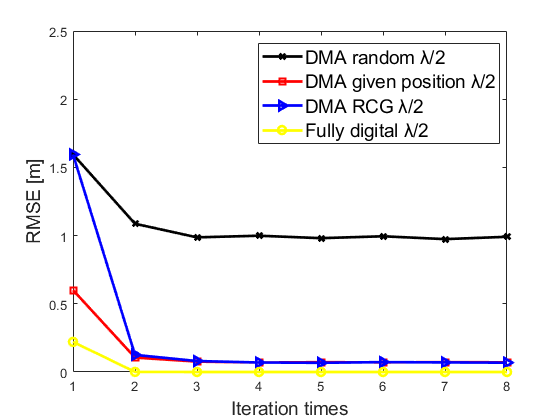} 
 \caption{Contrast of the number of iterations required for convergence via multiple contrast schemes. $N_{\rm d}=10$, $N_{\rm e}=50$, SNR is fixed as -5 dB, user position is set as $d_1 = 0.25 \, d_{\rm F}$ meters, $\theta_1 = \pi /6$, $d_2 = 0.25 \, d_{\rm F}$ meters, $\theta_2 = \pi /4$.
  %\textcolor{red}{FG: in the "x" axis, please write only "Number of RF chains"}
  } 
  \vspace{0.4cm}
  \label{fig9}
\end{figure}

To investigate the effectiveness of the $\bf Q$ design of the joint localization and DMA tuning algorithm in more detail, we investigated the number of iterations for the convergence of the Algorithm~\ref{alg:Alternating}, as shown in Fig.~\ref{fig9}, wherein the RMSE is the average of that of two users. To illustrate the convergence process in detail, we compared the convergence process between the random scheme and the RCG schemes. In Fig.~\ref{fig9}, the ``DMA RCG" denotes localization by Algorithm~\ref{alg:Alternating}, that obtaining the DMA tuning via the RCG-based method without prior knowledge of user position, and the ``DMA given position" denotes directly localization by Algorithm~\ref{alg:AP}, using the RCG-based method to get the DMA tuning which obtained from \eqref{max_prop_3} ignoring the the issue of unknown users location. It can be seen that the RMSE of the DMA RCG scheme converges from the random scheme to the same RMSE as the DMA-given position scheme. In addition, the speed of convergence depends on the accuracy of the initial position estimation, and the accuracy of the initial position estimation is related to the performance of the localization scheme thus the fully digital array always converges the fastest, while the DMA optimal scheme could achieve a closed converge speed to the fully digital. 
The alternating scheme generally requires more convergence times than the DMA opt-prior scheme, not only due to the initial position estimation but also because of its additional update for matrix $\bf Q$. Therefore, the DMA opt scheme compromises the DMA opt-prior scheme and the random scheme. However, it can be seen that the iteration of matrix $\bf Q$ is also efficient; only one additional iteration at most is required to converge to the optimal.

\begin{figure}
\centering 
\includegraphics[width=1\columnwidth]{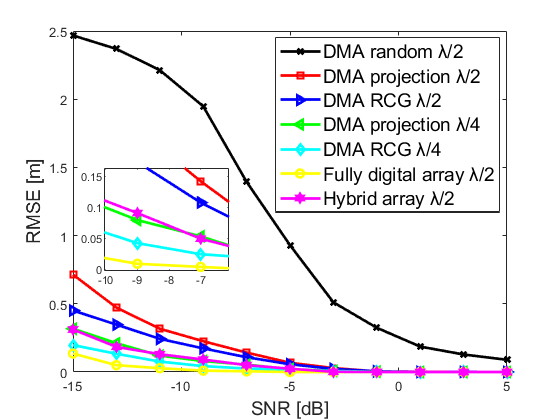}
\caption{Distortion contrast of multiple contrast schemes under different SNR with user position: $d_1 = 0.25 \, d_{\rm F}$ meters, $\theta_1 = \pi /6$; $d_2 = 0.25 \, d_{\rm F}$ meters, $\theta_2 = \pi /4$. $N_{\rm d}=10$, $N_{\rm e}$ is determined with the antenna space, $N_{\rm e}=50$ in $\lambda/2$ case, $N_{\rm e}=100$ in $\lambda/4$ case.
}
  \vspace{0.4cm}
\label{fig10}
\end{figure}

We show the effectiveness of the analog domain design scheme we proposed in Fig.~\ref{fig10}
%\FraCmt{Please write the units like SNR [dB]} 
which reports the RMSE of location estimation via different schemes under different SNRs. We considered a two-user localization scenario, where the caption describes the user's location. 
%For the DMA and hybrid array, we set the separation between microstrips as $5\lambda/2$, thus the number of microstrips in the DMA the number of sub-array in the hybrid array is $N_{\rm d} = 2L/(5\lambda) = 10$.
%, while in Fig.~\ref{fig:subfig:25RF}, we set the separation between microstrips as $\lambda$, thus the number of microstrips in this DMA is $N_{\rm d} = L/\lambda = 25$. 
The curve ``DMA RCG" and ``DMA projection" denote the DMA tuning scheme for localization using Algorithm~\ref{alg:Alternating} with optimization $\bf Q$ from projection-based solution and RCG-based solution, respectively. To indicate the performance improvement from tuning design, the DMA random scheme and the fully digital scheme are also provided for comparison, where the DMA random scheme is obtained by setting a set of random coefficients for $\bf Q$, while the fully digital scheme localizes the user with MLE method \cite{bibtex13} based on the fully digital array we just set.
As we can see in Fig.~\ref{fig10}, the DMA random scheme yields higher RMSE compared to the fully digital schemes, especially in the low SNR case. This is because the tunable coefficients matrix in the DMA random scheme represents only an interference effect for the MLE process. By tuning the DMA using Algorithm~\ref{alg:Alternating}, one significantly improves the RMSE, and the RCG-based scheme achieved better performance compared to the projection-based scheme; thus, in the next experiments, we use the RCG-base scheme as the representative of the DMA tuning scheme. Since without the Lorentzian constraint interference, the hybrid array scheme always outperforms DMA at the same number of antennas.
Meanwhile, the DMA tuning scheme with $\lambda/4$ spacing would further reduce RMSE by increasing the number of microstrip elements compared to the DMA tuning scheme with $\lambda/2$ spacing using the same tuning scheme. 

\begin{figure}
\centering 
\includegraphics[width=1\columnwidth]{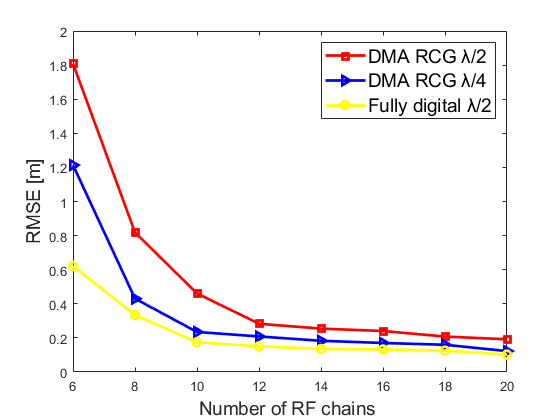}
\caption{Distortion contrast of multiple contrast schemes under different RF chain numbers. SNR = $-15$ dB, with user position: $d_1 = 0.25 \, d_{\rm F}$ meters, $\theta_1 = \pi /6$; $d_2 = 0.25 $ meters, $\theta_2 = \pi /4$. %$N_{\rm e}$ is determined with the antenna space, $N_{\rm e}=50$ in $\lambda/2$ case, $N_{\rm e}=100$ in $\lambda/4$ case.
  }
  \vspace{0.4cm}
\label{fig11}
\end{figure}

However, as shown in Fig.~\ref{fig10}, the RMSE of the DMA tuning scheme or hybrid array scheme with $\lambda/2$ antenna spacing is still higher than the fully digital schemes due to the reduction of signal dimension, even the DMA tuning scheme with $\lambda/4$ can only reduce this performance loss.
This is because the reduction of the RF chain can be seen as compression to the original received signal, resulting in inherent information loss. This also means that the increase in output signal dimension will alleviate the performance loss caused by compression behavior. In Fig.~\ref{fig11}, we discussed the impact of the number of RF chains on localization performance under the same DMA optimization, all arrays maintain the same aperture, while the arrangement density of microstrips controls the number of RF chains. As a comparison, the fully digital array is also provided, and in the same aperture, the RF chain number of the fully digital array is 50 times that of the DMA. With the increase in the number of RF chains, the performance of all schemes has improved as the antenna number increased. Among them, the DMA tuning scheme made more significant progress than the fully digital scheme, where the DMA scheme with $\lambda/4$ achieved a performance similar to that of the full-digital array scheme. This demonstrates our point and indicates that for DMA, the increased number of elements combined with their proper tuning via Algorithm~\ref{alg:Alternating} can achieve similar localization accuracy compared with fully digital arrays utilizing $\times 50$ more RF chains.

\section{Conclusions}
\label{sec:Conclusions}

In this work, we studied near-field multi-source position estimation based on the RF chain reduction array. We first presented the models for radiating near-field position estimation systems for the three different types of antenna architectures, including fully digital array, hybrid array, and DMA, where the latter two can be expressed in the same model formulation. We then formulated the optimization of the adjustable coefficients of the hybrid array and DMA to improve the accuracy of the position estimate and proposed an efficient algorithm for joint position estimate and adjustable coefficients design. Numerical results demonstrated that the design of adjustable coefficients of RF chain reduction array could significantly improve the near-field multi-source localization performance, reaching a fully digital implementation with at least one order of RF chain reduction. %In particular, the optimal coefficients design is shown as relevant to the positions of users, this coincides with the focusing effect in near-field communication.
%In this work, we studied near-field source position estimation based on a DMA array. We presented a model for DMA-based radiating near-field position estimation systems, and utilized a direct estimation method based on COA for localization. We then formulated the optimization of the DMA tuning to improve the accuracy of the position estimate, and proposed an efficient algorithm for joint position estimate and DMA tuning. Numerical results demonstrated that the DMA tuning design could significantly improve the near-field localization performance reaching and overtaking that of a fully digital implementation with at least one order of RF chains reduction. 

%and prove the effectiveness of the direct estimation method based on curvature-of-arrival in the DMA scenario. Our research shows that the receiving process of DMA can be regarded as the compression preprocessing of the received signal of all digital antennas, and our goal is to obtain the location information of the source from the compressed data. Therefore, we look for the phase design that minimizes the interference in the compression process, and we assess the effectiveness of the proposed approaches by simulations. 

%\ifFullVersion
%----------------------------------------------------------------------------------------
%	APPENDICES
%----------------------------------------------------------------------------------------
\vspace{-0.2cm}
\begin{appendices}
	\numberwithin{proposition}{section} 
	\numberwithin{lemma}{section} 
	\numberwithin{corollary}{section} 
	\numberwithin{remark}{section} 
	\numberwithin{equation}{section}

		\vspace{-0.2cm}
	\section{Proof of Theorem \ref{thm:B}}
	\label{app:Proof0}	
	Rewrite \eqref{max_prop} in scalar form:
	\begin{equation}
	\label{max2}
	\max _{{ q}_{i, l}} \sum_{i=1}^{N_{\rm d}} \sum_{m=1}^{M} \left|\sum_{l=1}^{N_{\rm e}} { q}_{i, l} { h}_{i, l} {\rm g}_{m,i, l} \right|^{2} , \quad \text { s.t. } {q}_{i, l} \in \mathcal{F}.
	\end{equation}
	\eqref{max2} can be decomposed into $N_{\rm d}$ sub-problems, where each sub-problem represents the optimization on the corresponding DMA microstrip. Substituting the expression of ${ g}_{m,i, l}$ in \eqref{receive} into \eqref{max2}, and letting ${ q}_{i, l}=e^{j \psi_{i, l}}$, we have the $i$-th sub-problem:
	\begin{equation}
	\label{max3}
    \begin{aligned}
	&\max _{\psi_{i, l}} \sum_{m=1}^{M} \left|\sum_{l=1}^{N_{\rm e}}  a_{m,i, l} e^{-\rho_{i, l}\alpha_{i}} e^{-j v_{m,i, l}} e^{-j\rho_{i, l} \beta_{i}} e^{j \psi_{i, l}} \right|^{2}. %\\&~~ \text { s.t. }\psi_{i,l} \in[0,2 \pi].
    \end{aligned}
	\end{equation}
 
    The optimal solution of \eqref{max2} can be obtained by solving sub-problem \eqref{max3} separately. It can be observed that the objective function of \eqref{max3} is composed by $M$ quadratic terms, where $m$-th term represents the phase adjustment of $i$-th microstrip on the received signal from $m$-th user, i.e., \eqref{max3} can be regarded as finding the global optimal phase design for all users.
    To this purpose, we firstly calculate the local optimal solution for the $m$-th user.
	Specifically, to maximize the $m$-th term of \eqref{max3}, according to the triangle inequality, we have 
	\begin{equation}
	\label{phase}
	\psi_{i, l}^{*(m)}=v_{m,i,l}+\rho_{i, l} \beta_{i}.
	\end{equation}
    The global optimal solution for \eqref{max3} should achieve a trade-off between all $\{ \psi_{i, l}^{*(m)} \}_{m = 1}^M$, i.e., it should achieves the minimum sum of Euclidean distances for all local optimal solutions, which can be calculated as
	%Observing \eqref{phase}, it is also the optimal phase design for the $m$-th user. Hence, problem \eqref{max3} can be regarded as an equilibrium problem, to obtain the global optimal solution of \eqref{max3}, we could find the centroid of $\left\{ \psi_{i, l}^{*(m)} \right\}_{m=1}^M$, which can be calculated as
	\begin{equation}
	\label{max4}
	\min _{\psi_{i, l}} {\rm C}(\psi_{i, l}) = \sum_{m=1}^{M} \left| \psi_{i, l} - \psi_{i, l}^{*(m)} \right|^{2}. 
	\end{equation}
	Obviously \eqref{max4} is convex, and the (optimal) solution can be found by solving the following equation
	%\FraCmt{Can you replace below ${\rm d}$ with $\partial$?}
	\begin{equation}
	\label{max5}
	\frac{\partial \left( {\rm C}(\psi_{i, l}) \right)}{\partial \left( \psi_{i, l} \right)} = 2M \psi_{i, l} - 2 \sum_{m=1}^{M} \psi_{i, l}^{*(m)} = 0. 
	\end{equation}
	whose solution is given by 
	\begin{equation}
	\label{max6}
	\psi_{i, l}^{*}= \frac{\sum_{m=1}^M v_{m,i,l}}{M}+\rho_{i, l} \beta_{i}. 
	\end{equation}
	This proves the theorem. 
	$\qed$
	%
	%-----------------------------------
	%	Proof of single user theorem
	%-----------------------------------
	
%	\vspace{-0.2cm}
%	\subsection{Proof of Theorem \ref{thm:SingleUser}}
%	\label{app:Proof1}	
%For a fixed weighting matrix ${\bf Q}$, the  digital precoding vector ${\bf w}$ that maximizes the objective function of \eqref{eq:optimization_problem_single} is 
% %
% \begin{equation} \label{eq: MRT}
%     {\bf w}^*=\sqrt{P_{\rm m}}\frac{\left({\bf a}^H\, \mathbf{H} \mathbf{Q}\right)^H}{\left\|{\bf a}^H\, \mathbf{H} \mathbf{Q}\right\|}.
% \end{equation}
% 
% The solution in \eqref{eq: MRT} implies that the maximal ratio transmission with the maximum available power is the optimal digital  precoding vector for any fixed ${\bf Q}$.
%
% 
%	
\end{appendices}

	%----------------------------------------------------------------------------------------
	%	BIBLIOGRAPHY
	%----------------------------------------------------------------------------------------
%	\newpage
	\bibliographystyle{IEEEtran}
	\bibliography{IEEEabrv,refs}

\begin{thebibliography}{10}
\providecommand{\url}[1]{#1}
\csname url@samestyle\endcsname
\providecommand{\newblock}{\relax}
\providecommand{\bibinfo}[2]{#2}
\providecommand{\BIBentrySTDinterwordspacing}{\spaceskip=0pt\relax}
\providecommand{\BIBentryALTinterwordstretchfactor}{4}
\providecommand{\BIBentryALTinterwordspacing}{\spaceskip=\fontdimen2\font plus
\BIBentryALTinterwordstretchfactor\fontdimen3\font minus \fontdimen4\font\relax}
\providecommand{\BIBforeignlanguage}[2]{{%
\expandafter\ifx\csname l@#1\endcsname\relax
\typeout{** WARNING: IEEEtran.bst: No hyphenation pattern has been}%
\typeout{** loaded for the language `#1'. Using the pattern for}%
\typeout{** the default language instead.}%
\else
\language=\csname l@#1\endcsname
\fi
#2}}
\providecommand{\BIBdecl}{\relax}
\BIBdecl

\bibitem{bibtex1}
C.~De~Lima \emph{et~al.}, ``Convergent communication, sensing and localization in 6{G} systems: An overview of technologies, opportunities and challenges,'' \emph{IEEE Access}, vol.~9, pp. 26\,902--26\,925, 2021.

\bibitem{wang2022location}
Z.~Wang \emph{et~al.}, ``Location awareness in beyond {5G} networks via reconfigurable intelligent surfaces,'' \emph{{IEEE} J. Sel. Areas Commun.}, vol.~40, no.~7, pp. 2011--2025, 2022.

\bibitem{torcolacci2023holographic}
G.~Torcolacci \emph{et~al.}, ``Holographic imaging with {XL}-{MIMO} and {RIS}: Illumination and reflection design,'' \emph{IEEE J. Sel. Topics Signal Process.}, To Appear, 2024.

\bibitem{bellini2024multi}
D.~T. Bellini \emph{et~al.}, ``Multi-view near-field imaging in nlos with non-reconfigurable em skins,'' \emph{arXiv preprint arXiv:2401.06891}, 2024.

\bibitem{nepa2017near}
P.~Nepa and A.~Buffi, ``Near-field-focused microwave antennas: Near-field shaping and implementation.'' \emph{{IEEE} Antennas Propag. Mag.}, vol.~59, no.~3, pp. 42--53, 2017.

\bibitem{zhang20226g}
H.~Zhang \emph{et~al.}, ``{6G} wireless communications: From far-field beam steering to near-field beam focusing,'' \emph{{IEEE} Commun. Mag.}, vol.~61, no.~4, pp. 72--77, 2023.

\bibitem{ahmed2018survey}
I.~Ahmed \emph{et~al.}, ``A survey on hybrid beamforming techniques in {5G}: Architecture and system model perspectives,'' \emph{{IEEE} Commun. Surveys Tuts.}, vol.~20, no.~4, pp. 3060--3097, 2018.

\bibitem{mendez2016hybrid}
R.~M{\'e}ndez-Rial \emph{et~al.}, ``Hybrid {MIMO} architectures for millimeter wave communications: Phase shifters or switches?'' \emph{IEEE Access}, vol.~4, pp. 247--267, 2016.

\bibitem{bibtex3}
L.~Taponecco, A.~D'Amico, and U.~Mengali, ``Joint {TOA} and {AOA} estimation for {UWB} localization applications,'' \emph{IEEE Trans. Wireless Commun.}, vol.~10, no.~7, pp. 2207--2217, 2011.

\bibitem{bibtex19}
D.~Dardari, P.~Closas, and P.~M. Djurić, ``Indoor tracking: Theory, methods, and technologies,'' \emph{IEEE Trans. Veh. Tech.}, vol.~64, no.~4, pp. 1263--1278, 2015.

\bibitem{bibtex5}
A.~Guerra, F.~Guidi, and D.~Dardari, ``Single-anchor localization and orientation performance limits using massive arrays: {MIMO} vs. beamforming,'' \emph{IEEE Trans. Wireless Commun.}, vol.~17, no.~8, pp. 5241--5255, 2018.

\bibitem{bibtex2}
A.~Elzanaty \emph{et~al.}, ``Towards 6{G} holographic localization: Enabling technologies and perspectives,'' \emph{IEEE Internet Things Mag.}, vol.~3, no.~6, pp. 138--143, 2023.

\bibitem{DarDecGueGui:J22}
D.~Dardari \emph{et~al.}, ``{LOS/NLOS} near-field localization with a large reconfigurable intelligent surface,'' \emph{{IEEE} Trans. Wireless Commun.}, vol.~21, no.~6, pp. 4282--4294, 2022.

\bibitem{9781656}
M.~Z. Win \emph{et~al.}, ``Location awareness via intelligent surfaces: {A} path toward holographic {NLN},'' \emph{IEEE Veh. Technol. Mag.}, vol.~17, no.~2, pp. 37--45, 2022.

\bibitem{isac}
N.~González-Prelcic \emph{et~al.}, ``The integrated sensing and communication revolution for 6g: Vision, techniques, and applications,'' \emph{Proceedings of the IEEE}, pp. 1--0, 2024.

\bibitem{bibtex6}
F.~Guidi and D.~Dardari, ``Radio positioning with {EM} processing of the spherical wavefront,'' \emph{IEEE Trans. Wireless Commun.}, vol.~20, no.~6, pp. 3571--3586, 2021.

\bibitem{bibtex9}
Y.-D. Huang and M.~Barkat, ``Near-field multiple source localization by passive sensor array,'' \emph{{IEEE} Trans. Antennas Propag.}, vol.~39, no.~7, pp. 968--975, 1991.

\bibitem{bibtex10}
B.~Ferguson and R.~Wyber, ``Wavefront curvature passive ranging in a temporally varying sound propagation medium,'' in \emph{Proc. MTS/IEEE Oceans, An Ocean Odyssey}, vol.~4, 2001, pp. 2359--2365 vol.4.

\bibitem{CRB}
L.~Khamidullina, I.~Podkurkov, and M.~Haardt, ``Conditional and unconditional {Cramér-Rao} bounds for near-field localization in bistatic {MIMO} radar systems,'' \emph{{IEEE} Trans. Signal Process.}, vol.~69, pp. 3220--3234, 2021.

\bibitem{9950340}
A.~A. D'Amico \emph{et~al.}, ``Cramér-rao bounds for holographic positioning,'' \emph{IEEE Trans. Signal Process.}, vol.~70, pp. 5518--5532, 2022.

\bibitem{bibtex11}
A.~Elzanaty \emph{et~al.}, ``Reconfigurable intelligent surfaces for localization: Position and orientation error bounds,'' \emph{IEEE Trans. Signal Process.}, vol.~69, pp. 5386--5402, 2021.

\bibitem{hybridRIS}
X.~Zhang and H.~Zhang, ``Hybrid reconfigurable intelligent surfaces-assisted near-field localization,'' \emph{{IEEE} Commun. Lett.}, vol.~27, no.~1, pp. 135--139, 2023.

\bibitem{HolographicRIS}
X.~Gan \emph{et~al.}, ``Near-field localization for holographic {RIS} assisted mmwave systems,'' \emph{{IEEE} Commun. Lett.}, vol.~27, no.~1, pp. 140--144, 2023.

\bibitem{bibtex13}
A.~Guerra \emph{et~al.}, ``Near-field tracking with large antenna arrays: Fundamental limits and practical algorithms,'' \emph{IEEE Trans. Signal Process.}, vol.~69, pp. 5723--5738, 2021.

\bibitem{2Dmusic}
X.~Zhang \emph{et~al.}, ``Localization of near-field sources: A reduced-dimension {MUSIC} algorithm,'' \emph{{IEEE} Commun. Lett.}, vol.~22, no.~7, pp. 1422--1425, 2018.

\bibitem{MLE}
C.~Cheng \emph{et~al.}, ``An efficient maximum-likelihood-like algorithm for near-field coherent source localization,'' \emph{{IEEE} Trans. Antennas Propag.}, vol.~70, no.~7, pp. 6111--6116, 2022.

\bibitem{rahal2023performance}
M.~Rahal \emph{et~al.}, ``Performance of {RIS}-aided near-field localization under beams approximation from real hardware characterization,'' \emph{EURASIP J. Wireless Commun. Netw.}, vol. 2023, no.~1, pp. 1--23, 2023.

\bibitem{MUSIC}
K.~Hu, S.~P. Chepuri, and G.~Leus, ``Near-field source localization: Sparse recovery techniques and grid matching,'' in \emph{IEEE Sensor Array and Multichannel Signal Processing Workshop (SAM)}, 2014, pp. 369--372.

\bibitem{3Dmusic}
Z.~Li \emph{et~al.}, ``3-{D} localization for near-field and strictly noncircular sources via centro-symmetric cross array,'' \emph{IEEE Sensors Journal}, vol.~21, no.~6, pp. 8432--8440, 2021.

\bibitem{zirtiloglu2022power}
T.~Zirtiloglu \emph{et~al.}, ``Power-efficient hybrid {MIMO} reciever with task-specific beamforming using low-resolution {ADCs},'' in \emph{Proc. IEEE ICASSP}, 2022.

\bibitem{ZenZha:J16}
Y.~Zeng and R.~Zhang, ``Millimeter wave {MIMO} with lens antenna array: A new path division multiplexing paradigm,'' \emph{IEEE Trans. Commun.}, vol.~64, no.~4, pp. 1557--1571, 2016.

\bibitem{abu2021near}
Z.~Abu-Shaban \emph{et~al.}, ``Near-field localization with a reconfigurable intelligent surface acting as lens,'' in \emph{Proc. IEEE Int. Conf. on Commun. (ICC)}, 2021.

\bibitem{levy2024rapid}
O.~Levy and N.~Shlezinger, ``Rapid hybrid modular receive beamforming via learned optimization,'' in \emph{IEEE International Conference on Acoustics, Speech and Signal Processing (ICASSP)}, 2024, pp. 12\,826--12\,830.

\bibitem{gong2019rf}
T.~Gong \emph{et~al.}, ``{RF} chain reduction for {MIMO} systems: A hardware prototype,'' \emph{{IEEE} Syst. J.}, vol.~14, no.~4, pp. 5296--5307, 2020.

\bibitem{lavi2023learn}
O.~Lavi and N.~Shlezinger, ``Learn to rapidly and robustly optimize hybrid precoding,'' \emph{{IEEE} Trans. Commun.}, vol.~71, no.~10, pp. 5814--5830, 2023.

\bibitem{bibtex20}
N.~Shlezinger \emph{et~al.}, ``Dynamic metasurface antennas for 6{G} extreme massive {MIMO} communications,'' \emph{IEEE Wireless Commun.}, vol.~28, no.~2, pp. 106--113, 2021.

\bibitem{shlezinger2019dynamic}
------, ``Dynamic metasurface antennas for uplink massive {MIMO} systems,'' \emph{{IEEE} Trans. Commun.}, vol.~67, no.~10, pp. 6829--6843, 2019.

\bibitem{bibtex15}
H.~Zhang \emph{et~al.}, ``Beam focusing for near-field multi-user {MIMO} communications,'' \emph{{IEEE} Trans. Wireless Commun.}, vol.~21, no.~9, pp. 7476--7490, 2022.

\bibitem{DMAlocation}
Q.~Yang \emph{et~al.}, ``Near-field localization with dynamic metasurface antennas,'' in \emph{IEEE International Conference on Acoustics, Speech and Signal Processing (ICASSP)}, 2023.

\bibitem{DMAfocusing}
P.~Gavriilidis and G.~C. Alexandropoulos, ``Near-field beam tracking with extremely large dynamic metasurface antennas,'' \emph{arXiv preprint arXiv:2406.01488}, 2024.

\bibitem{digital}
J.~S. Herd and M.~D. Conway, ``The evolution to modern phased array architectures,'' \emph{Proc. {IEEE}}, vol. 104, no.~3, pp. 519--529, 2016.

\bibitem{ioushua2019family}
S.~S. Ioushua and Y.~C. Eldar, ``A family of hybrid analog--digital beamforming methods for massive {MIMO} systems,'' \emph{{IEEE} Trans. Signal Process.}, vol.~67, no.~12, pp. 3243--3257, 2019.

\bibitem{bibtex21}
T.~Sleasman \emph{et~al.}, ``Waveguide-fed tunable metamaterial element for dynamic apertures,'' \emph{{IEEE} Antennas Wireless Propag. Lett.}, vol.~15, pp. 606--609, 2016.

\bibitem{bibtex14}
C.~L. Holloway \emph{et~al.}, ``An overview of the theory and applications of metasurfaces: The two-dimensional equivalents of metamaterials,'' \emph{IEEE Antennas Propag. Mag.}, vol.~54, no.~2, pp. 10--35, 2012.

\bibitem{source-numb}
L.~Huang \emph{et~al.}, ``{MMSE-Based} {MDL} method for robust estimation of number of sources without eigendecomposition,'' \emph{{IEEE} Trans. Signal Process.}, vol.~57, no.~10, pp. 4135--4142, 2009.

\bibitem{alternating}
I.~Ziskind and M.~Wax, ``Maximum likelihood localization of multiple sources by alternating projection,'' \emph{{IEEE} Trans. Acoust., Speech, Signal Process.}, vol.~36, no.~10, pp. 1553--1560, 1988.

\bibitem{DMApower}
H.~Zhang \emph{et~al.}, ``Near-field wireless power transfer with dynamic metasurface antennas,'' in \emph{IEEE International Workshop on Signal Processing Advances in Wireless Communication (SPAWC)}, 2022.

\end{thebibliography}

\end{document}